\documentclass[]{aa}  
\usepackage{graphicx}
\usepackage{txfonts}
\usepackage{natbib}
\bibpunct{(}{)}{;}{a}{}{,} 	

\begin{document}

   \title{Three-Dimensional Hydrodynamical Simulations of Surface Convection
   in Red Giant Stars}

   \subtitle{Impact on spectral line formation and abundance analysis}
   \titlerunning{3D Hydrodynamical Simulations of Surface Convection
   in Red Giants}
   
   \author{R. Collet\inst{1,2}
          \and
          M. Asplund\inst{3}
	  \and
	  R. Trampedach\inst{3} 
	  } 

   \offprints{R. Collet}

   \institute{ Max-Planck-Institut f{\"u}r Astrophysik, Karl-Schwarzschild-Str.~1, 
	       Postfach 1317, D--85741 Garching b. M{\"u}nchen, Germany \\
	       \email{remo@mpa-garching.mpg.de}
	 \and
		Department of Astronomy and Space Physics,
        	Uppsala University, BOX 515, SE--751 20 Uppsala, Sweden \\
         \and
             	Research School of Astronomy \& Astrophysics,
        	Mount Stromlo Observatory, Cotter Road, Weston ACT 2611, Australia\\
             	\email{martin@mso.anu.edu.au, art@mso.anu.edu.au} 
             }

   \date{(Accepted for publication on \aap)}

  \abstract
   {} 
   {We investigate the impact of realistic three-dimensional (3D) hydrodynamical model atmospheres
   of red giant stars at different metallicities on the formation of spectral lines 
   of a number of ions and molecules. }
   {We carry out realistic, ab initio, 3D, hydrodynamical simulations of 
    surface convection at the surface of red giant stars with 
    varying effective temperatures and metallicities. 
    We use the convection simulations as time-dependent
    hydrodynamical model stellar atmospheres to calculate spectral lines 
    for a number of atomic 
    (\ion{Li}{i}, \ion{O}{i}, \ion{Na}{i}, \ion{Mg}{i}, \ion{Ca}{i}, 
    \ion{Fe}{i}, and \ion{Fe}{ii}) and molecular (CH, NH, and OH) lines
    under the assumption of local thermodynamic equilibrium (LTE). 
    We carry out a differential comparison of the line strengths computed in 3D 
    with the results of analogous line formation calculations for classical, 
    1D, hydrostatic, plane-parallel \textsc{marcs} model atmospheres in order
    to estimate the impact of 3D models on the derivation of elemental abundances.} 
   {The temperature and density inhomogeneities and correlated velocity fields 
    in 3D models, as well as the differences between the mean 3D stratifications 
    and corresponding 1D model atmospheres significantly affect the
    predicted strengths of spectral lines.
    Under the assumption of LTE,
    the low atmospheric temperatures encountered in 3D model
    atmospheres of very metal-poor giant stars cause spectral lines 
    from neutral species and molecules to appear stronger than within 
    the framework of 1D models.
    As a consequence, elemental abundances derived from these lines using 3D models
    are significantly lower than according to 1D analyses.
    In particular, the differences between 3D and 1D abundances of C, N, and O derived from CH, NH, and
    OH weak low-excitation lines are found to be in the range $-0.5$~dex to $-1.0$~dex for the
    the red giant stars at $[\mathrm{Fe/H}]=-3$ considered here. 
    At this metallicity, large negative corrections (about $-0.8$~dex) are also found, in LTE, for 
    weak low-excitation \ion{Fe}{i} lines.  
    We caution, however, that the neglected departures from LTE might be significant for
    these and other elements and comparable to the effects due to stellar granulation.}
   {}


   \maketitle
\section{Introduction}
Stars in the red giant phase of their evolution are characterized
by large stellar diameters and increased luminosities compared with unevolved objects.
Because of their high intrinsic brightnesses, red giants represent 
natural targets for a wide range of observational studies.
Stars of this luminosity class are especially suitable for investigations of 
distant stellar systems in the Milky Way as well as in
other galaxies of the Local Group.
Giant stars are extensively used in spectroscopic analyses
for tracing elemental abundances in distant stellar populations.
A significant fraction of the halo and disk stars included in various large-scale stellar
abundance analyses  \citep[e.g.][]{mcwilliam95b, ryan96,fulbright00}
are indeed giants.
More recently, the ESO ``First Stars'' programme 
\citep[e.g.][]{hill02,cayrel04,spite05}
has provided a systematic and homogeneous study of the chemical
compositions of a large sample of extremely metal-poor giant as well as dwarf stars 
($\mathrm{[Fe/H]}\la-2.5$) from the HK survey by \citet{beers92,beers99}.
The recently discovered extreme halo star HE\,$0107-5240$ \citep{christlieb02},
one of the most iron-poor star known to date, is also a giant.
Red giants are also used to study Galactic 
metallicity gradients in the Galactic disk and open clusters
\citep{yong05,carney05} and star-to-star elemental abundance
variations within globular clusters \citep[for a review, see ][]{gratton04}.
The results of these and similar observational studies provide valuable information
about fundamental astrophysical processes and are essential for our
understanding of stellar nucleosynthesis and internal mixing
as well as Galactic chemical evolution.
Accurate determinations of elemental abundances in metal-poor stars
and red giants in particular are crucial for deducing the physical and
evolutionary properties of the very first generation of stars \citep{weiss04,iwamoto05,meynet06}
and distinguishing among the various proposed scenarios of
chemical enrichment of the Galaxy.

As for other late-type stars, ordinary classical abundance analyses 
of red giants involve theoretical 1D model atmospheres constructed
under the assumptions of plane-parallel geometry (or
spherical symmetry), hydrostatic equilibrium, and flux constancy
\citep[e.g.][]{gustafsson94}.
The simplifying approximation of LTE is also normally adopted. 
Furthermore, 1D modelling of stellar atmospheres generally relies on
a rudimentary treatment of convective energy transport,
such as the mixing length theory \citep{boehm-vitense58} or 
one of its derivatives \citep[e.g.][]{canuto91}, all dependent on a 
number of tunable but not necessarily physically well motivated free parameters.
In late-type stars, however, the convective zone 
reaches and significantly affects the regions from which the
stellar flux is emitted.
High spatial resolution imaging of the solar photosphere \citep[e.g.][]{title90,spruit90,carlsson04}
reveals that the surface of the Sun is dominated by a distinct granulation
pattern reflecting the bulk flows in the convective zone deeper inside.
 Observational diagnostics of stellar granulation in general are more limited,
as the surfaces of most other stars cannot be directly resolved.
Nonetheless, other distinguishing signatures (e.g. wavelength 
shifts and asymmetries of line profiles) of  the presence of photospheric 
velocity fields and correlated temperature inhomogeneities
can be identified in high-resolution spectra of late-type stars
\citep{dravins82,dravins87b,dravins87a,allende02a}.
These observable manifestations of stellar surface convection immediately reveal
that the assumptions of hydrostatic equilibrium and flux constancy 
in 1D model atmospheres of late-type stars are questionable and might be
inadequate for high-precision stellar abundance analyses.
Furthermore, 1D models can neither predict the
strengths and shapes of lines without invoking the ad-hoc fudge factors
micro- and macro-turbulence to account for non-thermal Doppler 
broadening of spectral lines associated with bulk flows in the
stellar atmosphere.	
During the past three decades, realistic 3D hydrodynamical
simulations of stellar surface convection have become feasible thanks to
the advances in computer technology and the development of efficient 
numerical algorithms 
\citep[e.g.][]{nordlund82,nordlund90,stein98,asplund99,freytag02,ludwig02,carlsson04,voegler04}.
At present, 3D hydrodynamical simulations of surface convection 
have primarily been carried out for main sequence stars and subgiants of spectral types A to M at
different metallicities; due to the disparate time scales for 
radiation transport and convection, convection simulations of red giant stars
appear in general more computationally demanding, but they are currently being developed
\citep{collet06, kucinskas06}.

 Three-dimensional simulations of stellar surface convection can be used as
time-series of hydrodynamical model atmospheres to study the effects of photospheric
inhomogeneities and velocity fields on the formation of spectral lines.
Recent analyses based on state-of-the-art 3D time-dependent 
simulations of surface convection in the Sun, dwarfs and subgiants
\citep[e.g.][ and references therein]{asplund99, asplund01, asplund05} 
indicate that the structural differences between 3D hydrodynamical and 1D hydrostatic 
model stellar atmospheres can have a significant impact on the predicted strengths of
synthetic spectral lines and hence lead to severe systematic
effects on the derivation of elemental abundances,
in particular at low metallicity.
In the present paper, following the spirit of these previous works, we carry out the first
3D hydrodynamical simulations of surface convection in red giant stars
with metallicities ranging from solar down to $[\mathrm{Fe/H}]=-3$.
We examine the main differences between the structures of the
3D hydrodynamical simulations and 1D hydrostatic model atmospheres
computed for the same stellar parameters and study
their effect on spectral line formation and abundance analyses of giant stars.

\section{Methods}
\subsection{Convection simulations}
\label{sec:convec-sim}

\begin{table}
\begin{minipage}[t]{\columnwidth}
\caption{Details of the 3D hydrodynamical simulations.}
\label{tab:param}
\centering
\renewcommand{\footnoterule}{}  
\begin{tabular}{ccccc}
\hline\hline 
$\langle T_\mathrm{eff}\rangle$~\footnote{Temporal 
average and  standard deviation of the emergent effective temperatures.}&
$\log{g}$ & 
$[\mathrm{Fe/H}]$ & 
$x$,$y$,$z$-dimensions & 
time~\footnote{Time span of the parts of simulations used for spectral line formation purposes.} \\   
     $\mathrm{[K]}$	& $[\mathrm{cgs}]$ &		& $\mathrm{[Mm]}$                   	& $\mathrm{[min]}$ \\ 
\hline
$4697{\pm}18$	& $2.2$	&	$+0.0$	& $1250{\times}1250{\times}610$	& 13\,000	\\
$4717{\pm}12$	& $2.2$	&	$-1.0$	& $1125{\times}1125{\times}415$	&~~6\,000	\\
$4732{\pm}~8~$	& $2.2$	&	$-2.0$	& $1065{\times}1065{\times}360$	&~~8\,000	\\
$4858{\pm}10$	& $2.2$	&	$-3.0$	& $1065{\times}1065{\times}360$	& 13\,500	\\
\noalign{\smallskip}
$4983{\pm}36$	& $2.2$	&	$+0.0$	& $1290{\times}1290{\times}960$	& 15\,000	\\
$5131{\pm}19$	& $2.2$	&	$-1.0$	& $1250{\times}1250{\times}540$	& 11\,000	\\
$5035{\pm}13$	& $2.2$	&	$-2.0$	& $1150{\times}1150{\times}430$	& 10\,500	\\
$5128{\pm}10$	& $2.2$	&	$-3.0$	& $1150{\times}1150{\times}430$	&~~8\,000	\\
\hline 
\end{tabular}
\end{minipage}
\end{table}

We use the 3D, time-dependent, compressible, explicit, 
radiative-hydrodynamical code by \citet{stein98} 
to construct sequences of surface convection simulations of  
red giant stars with varying effective temperatures and metallicities.
The hydrodynamical equations of mass, momentum, and energy
conservation are solved together with the 3D radiative
transfer equation on a Eulerian mesh with $100{\times}100{\times}125$
grid points. 
The physical domains of the simulations are set
large enough to cover about ten granules simultaneously in the horizontal plane
and eleven pressure scales in the vertical direction.
In terms of continuum optical depth at $\lambda\!=\!5000$~{\AA}, the
simulations extend from $\log\tau_{5000} \la -5$ down to 
$\log\tau_{5000} \ga 7$.
The depth scales are optimized to provide the highest spatial
resolution in those layers near the optical
surface where the vertical temperature gradients are steepest.
For the simulations, we employ open boundaries vertically and
periodic boundaries horizontally.
At each time step the radiative transfer equation is solved along one 
vertical and eight inclined rays;
the opacities are grouped into four opacity bins \citep{nordlund82}
and local thermodynamic equilibrium (LTE) without continuous 
scattering terms in the source function ($S_\nu\!=\!B_\nu$)
is assumed throughout the calculations.
The adopted equation of state comes from \citet{mihalas88} and
accounts for the effects of ionization, excitation, and 
dissociation of 15 of the most abundant elements, as well as
the H$_2$ and H$_2^+$ molecules.
Continuous opacities come from the Uppsala opacity package 
\citep[][ and subsequent updates]{gustafsson75} and line opacity
data from \citet{kurucz92,kurucz93}.

For the present work, we have generated two series of
3D hydrodynamical surface convection simulations of 
red giant stars with surface gravity $\log{g}=2.2$~(cgs).
The first series comprises simulations of red giants with
 $T_\mathrm{eff}{\simeq}4750$~K and 
metallicities $[\mathrm{Fe/H}]=0$, $-1$, $-2$, and
$-3$ while the second series corresponds to red giants
in the same metallicity range but with somewhat higher effective temperatures
($T_\mathrm{eff}{\simeq}5050$~K).
For all hydrodynamical simulations, we adopt the solar chemical composition 
from \citet{grevesse98} with the abundances of all metals
scaled proportionally from solar to the relevant $[\mathrm{Fe/H}]$ 
and without $\alpha$-enhancements.
The initial snapshots for the present red giant simulations
are obtained from simulations performed at a lower numerical resolution
($50{\times}50{\times}125$), which have been running for
sufficiently long times to cover several convective
turn-over time-scales and allow thermal relaxation
to occur.
The initial snapshots of the lower resolution simulations
in turn have been constructed by appropriately scaling a set of
previous simulations of turn-off stars by \citet{asplund01} to
the new stellar parameters, using the experience from 
classical 1D, hydrostatic stellar models.

Some relevant parameters and characteristic physical
quantities of the red giant convection 
simulations are given in Table~\ref{tab:param}.
While the geometrical depths of the simulation domains
resulting from the scaling process vary depending on the metallicity,
all convection simulations span approximately five pressure 
scale heights below and six pressure scale heights above
the optical surface. 
We deem this sufficient to realistically simulate the velocity fields
and spatial structures at the stellar surface.
It is worthwhile mentioning that, in the convection
simulations, the entropy of the inflowing gas at the lower boundary
replaces the effective temperature
as a constant and independent input parameter \citep{stein98}.
Consequently, the emergent effective temperatures 
of the simulations vary slightly with time,
the total outgoing radiative flux being susceptible 
to the evolution of the surface granulation pattern.
The actual value of the temporally averaged $T_\mathrm{eff}$
ultimately depends on the entropy of the inflowing gas.
Constructing a simulation with a specific value of the temporal averaged
effective temperature requires a careful
fine-tuning of the entropy parameter.
As this procedure is very time-consuming, and as we are
primarily interested in a differential comparison between
3D and 1D models,
we consider it satisfactory for the scope
of the present paper to settle for values 
reasonably close to the targeted effective
temperatures for the two suites of convection simulations.

From a qualitative point of view, the atmospheric structures 
and gas flows resulting from the convection simulations
are fairly similar to the ones previously found
by \citet{asplund99} and \citet{asplund01} for dwarfs and turn-off stars. 
The warm isentropic gas ascending from the stellar interior 
rapidly cools as it approaches the optical surface, where it
also loses entropy; the cooled gas eventually turns over and falls back 
toward the interior due to negative buoyancy.
The morphology and the evolution of the resultant granulation patterns,
with warm, large upflows amidst cool, narrow downdrafts,
are essentially the same as for solar-type stars.

\begin{figure*}
   \centering
   \resizebox{\hsize}{!}{
	\includegraphics{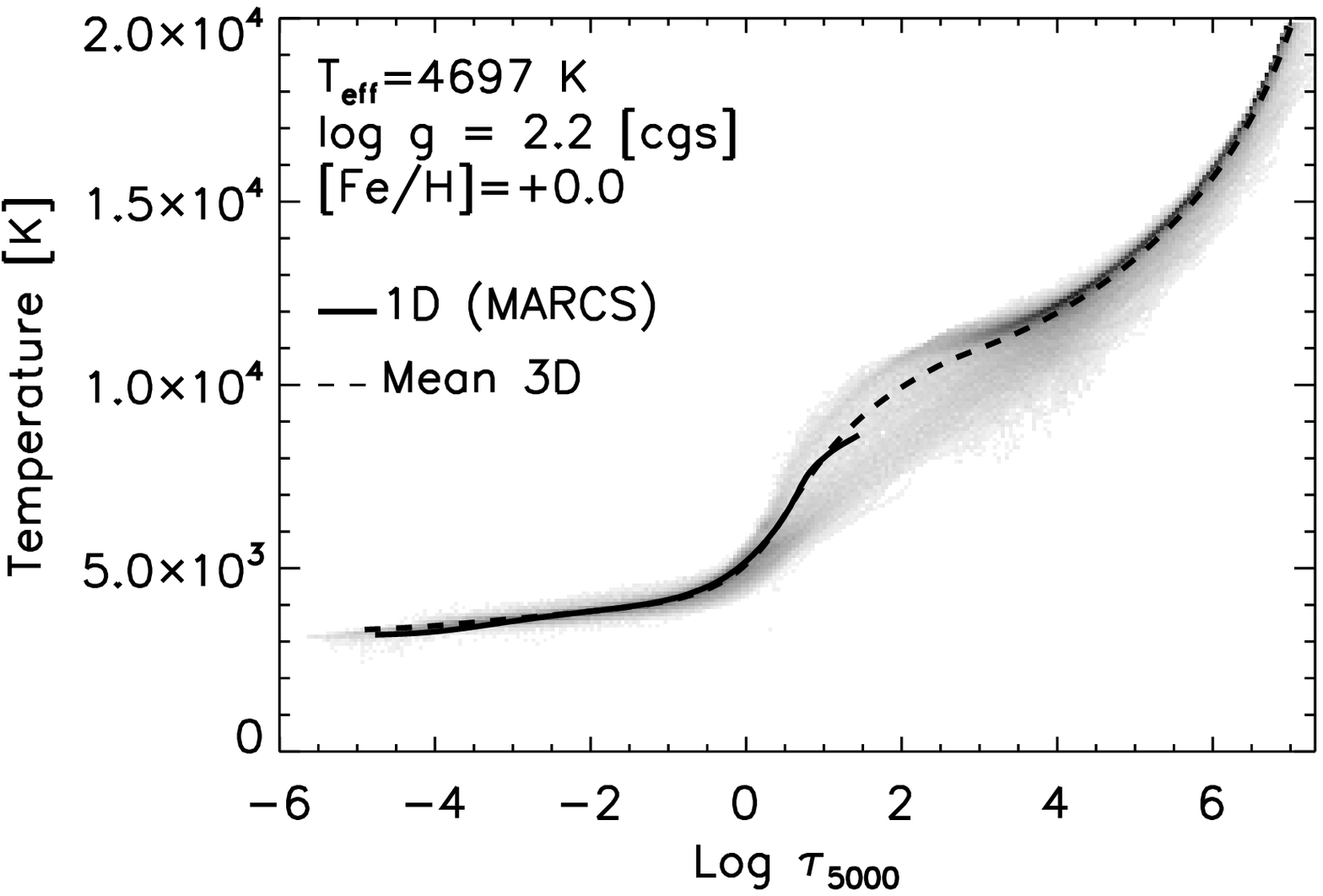}
	\includegraphics{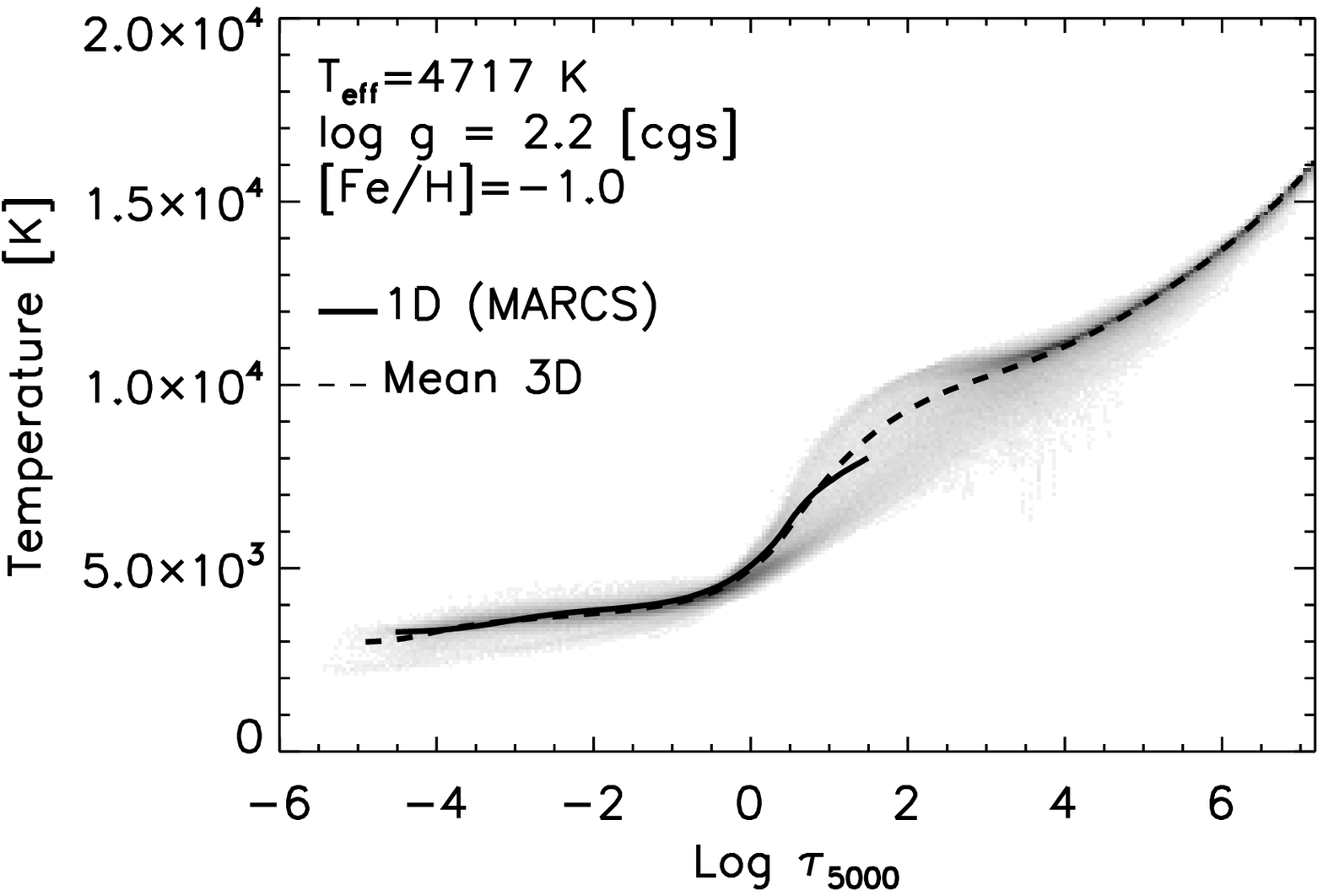} }
   \resizebox{\hsize}{!}{
	\includegraphics{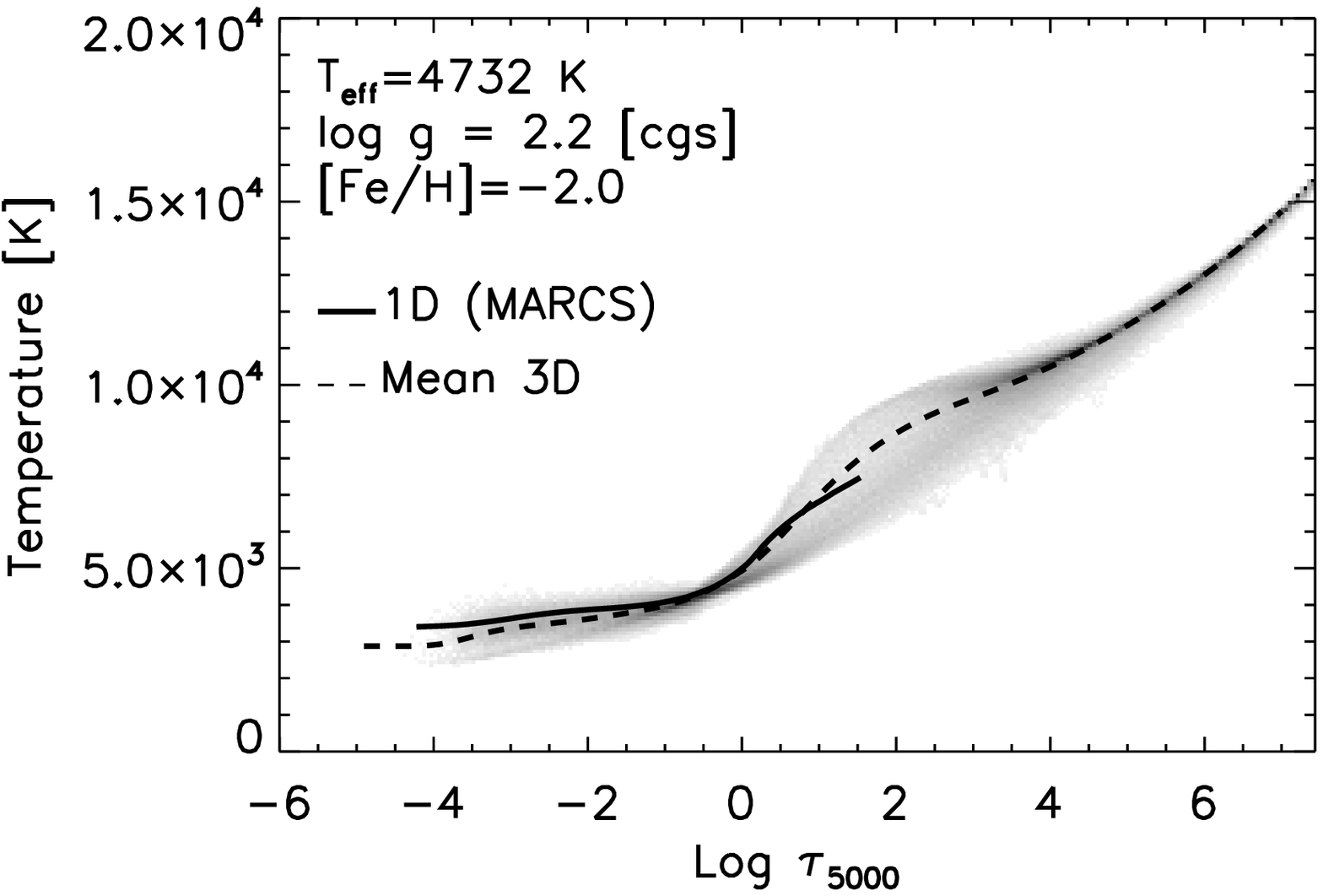}
	\includegraphics{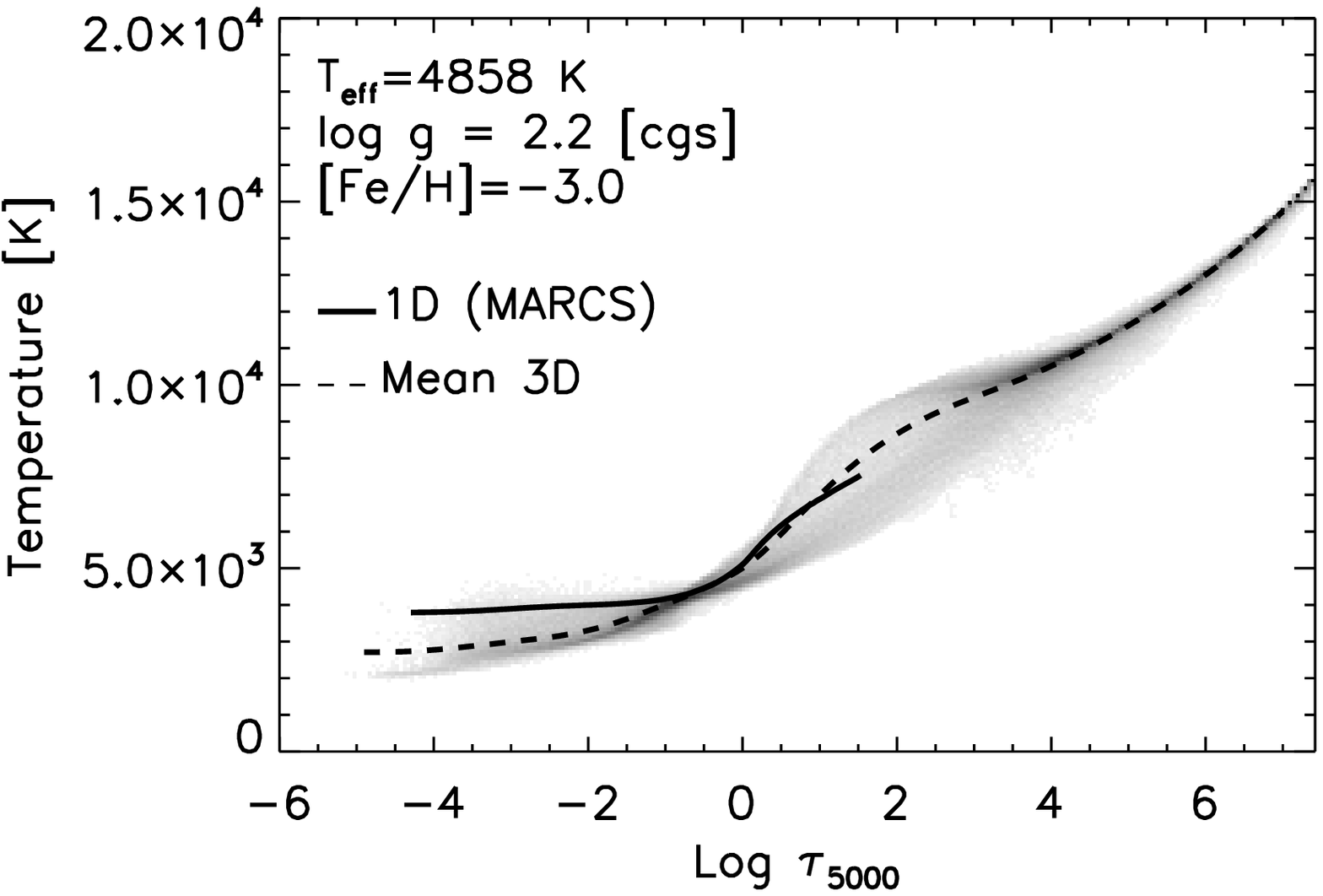} }
   \caption{Thermal structures of four snapshots of 3D 
   hydrodynamical simulations of red giants at different metallicities. 
   \emph{Gray area}: temperature distribution as a function of optical 
   depth at $\lambda=5000$~{\AA} in the  3D convection simulations. 
   Darker areas indicate temperature values with higher probability. 
   \emph{Thin dashed line}: Mean temperature stratifications of the 
   3D simulations (averaged over surfaces of equal optical depth at 
   $\lambda=5000$~{\AA}). \emph{Thick solid line}: Temperature stratifications 
   of the corresponding {\sc marcs} models.}
     \label{fig:models}
\end{figure*}

 \begin{figure*}
   \centering
  \resizebox{\hsize}{!}{
       \includegraphics{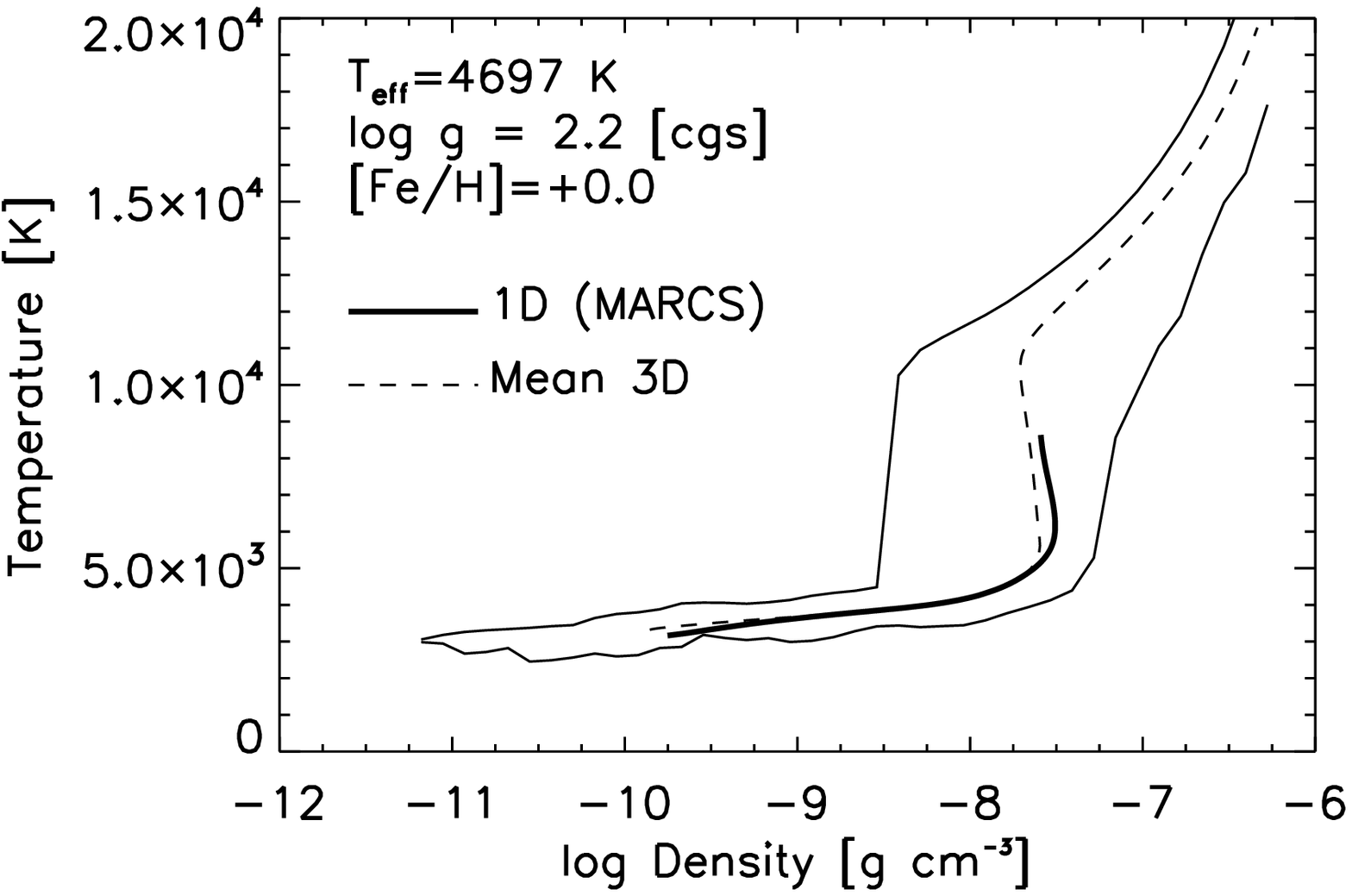}
       \includegraphics{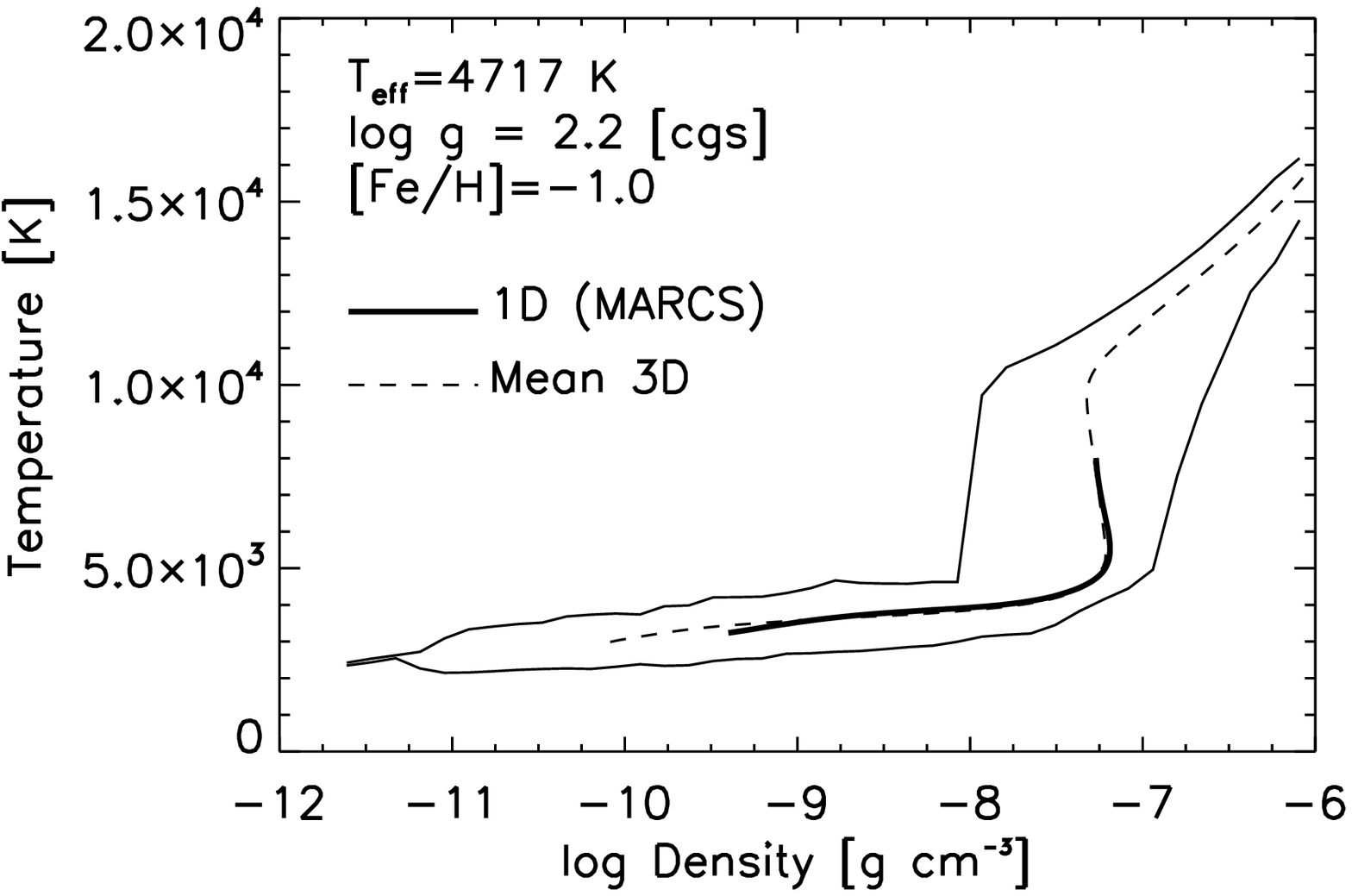} }
   \resizebox{\hsize}{!}{
       \includegraphics{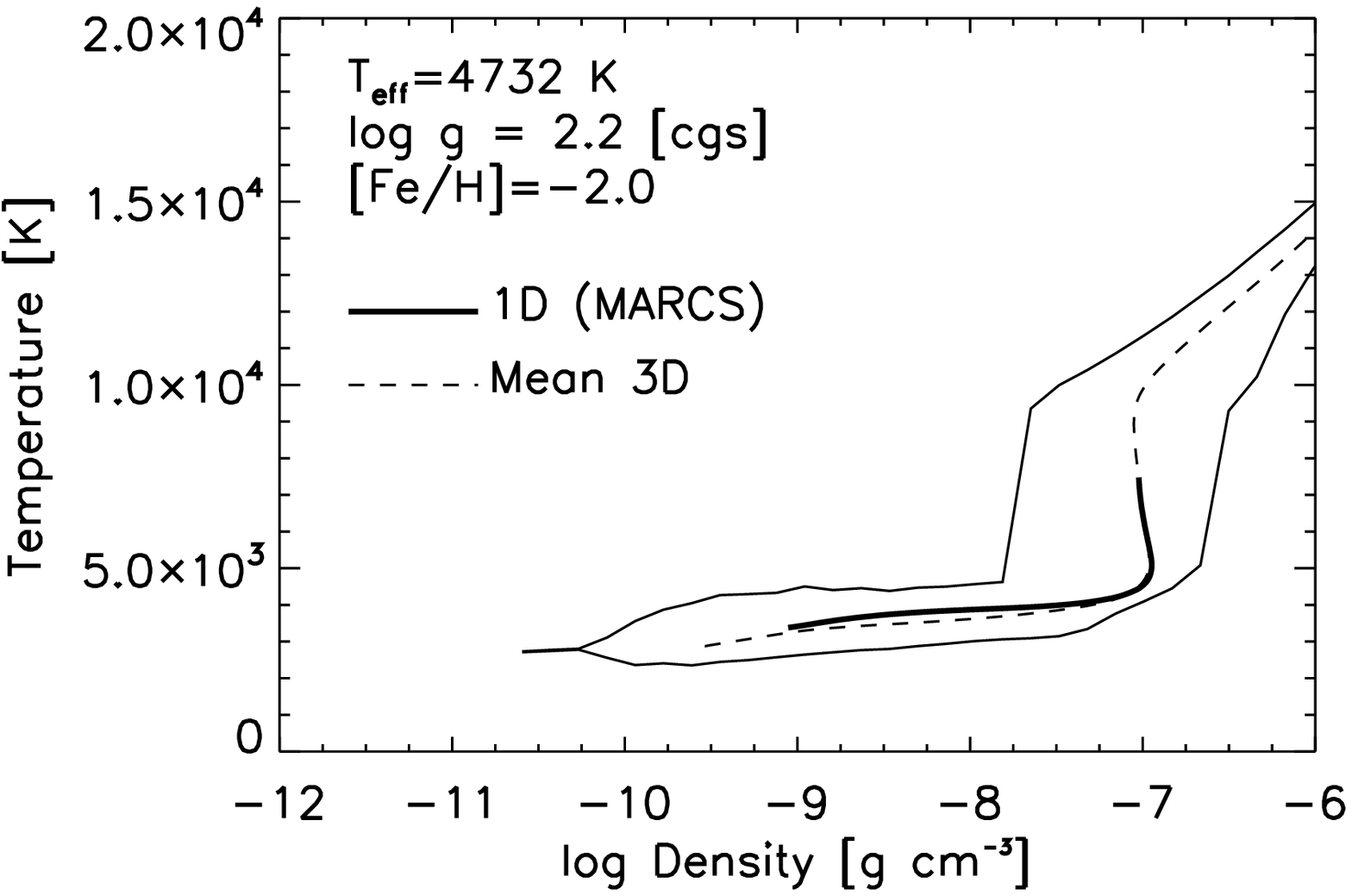}
       \includegraphics{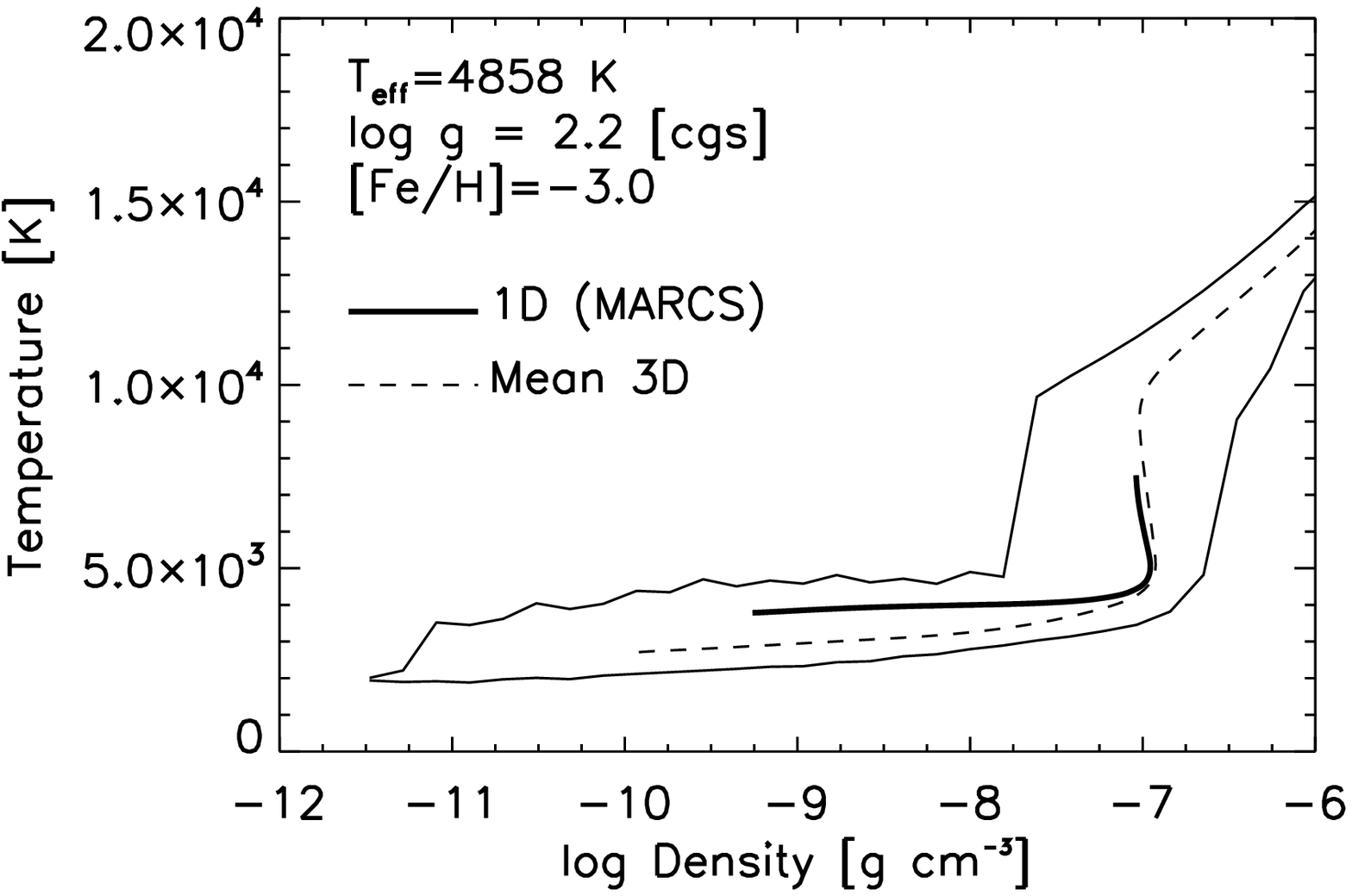} }
   \caption{Thermal structures of four snapshots of 3D hydrodynamical simulations 
   of red giants at different metallicities. \emph{Thin solid line}: extreme temperatures at
   a given density in the 3D convection simulation. \emph{Thin dashed line}: Mean
   temperature-density stratifications of the 3D simulations (averaged over
   surfaces of equal optical depth at $\lambda=5000$~{\AA}). \emph{Thick solid line}:
   Temperature-density stratification for the corresponding {\sc marcs} models.}
     \label{fig:models-rho}
\end{figure*}

 \begin{figure*}
   \centering
  \resizebox{\hsize}{!}{
       \includegraphics{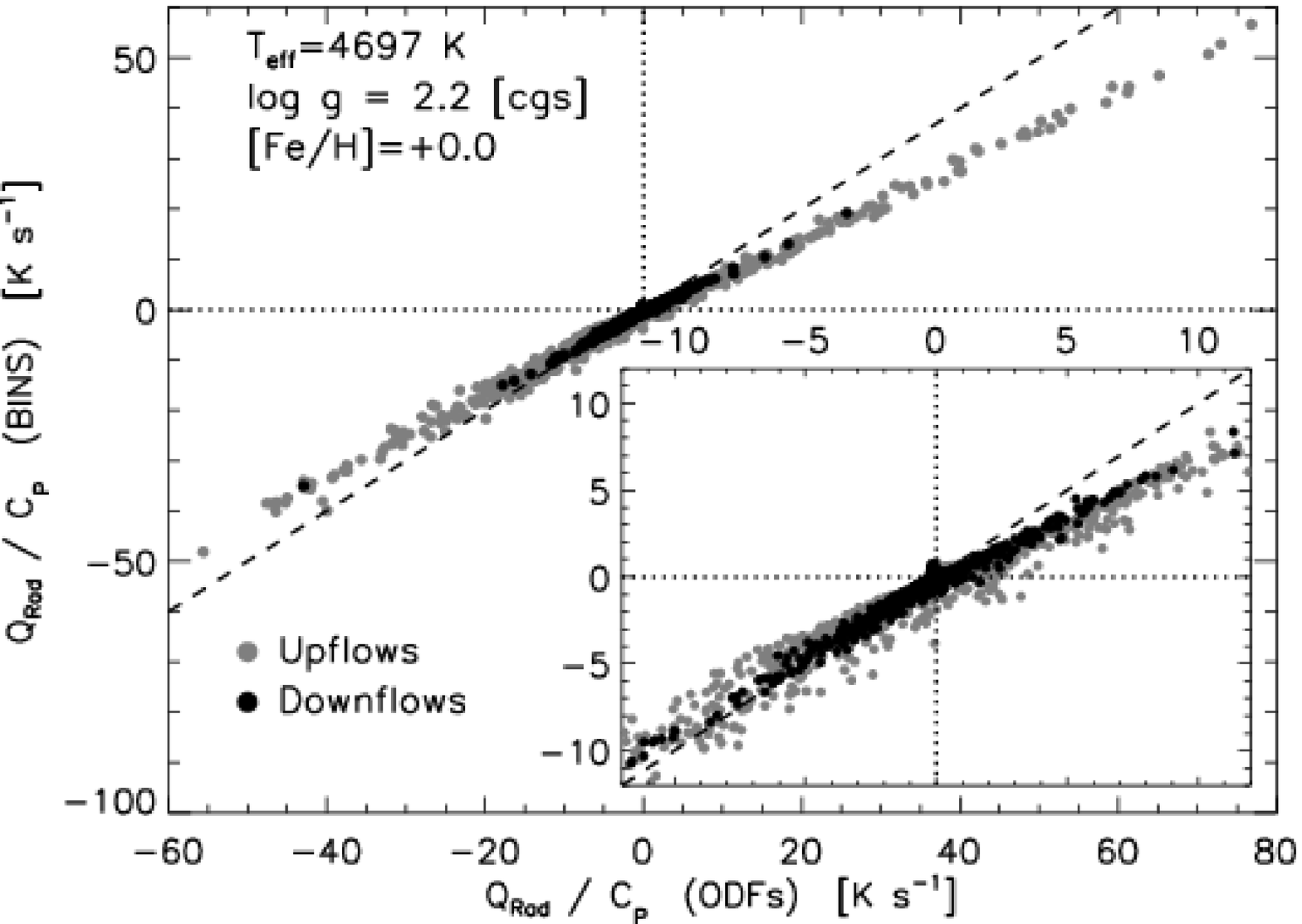}
       \includegraphics{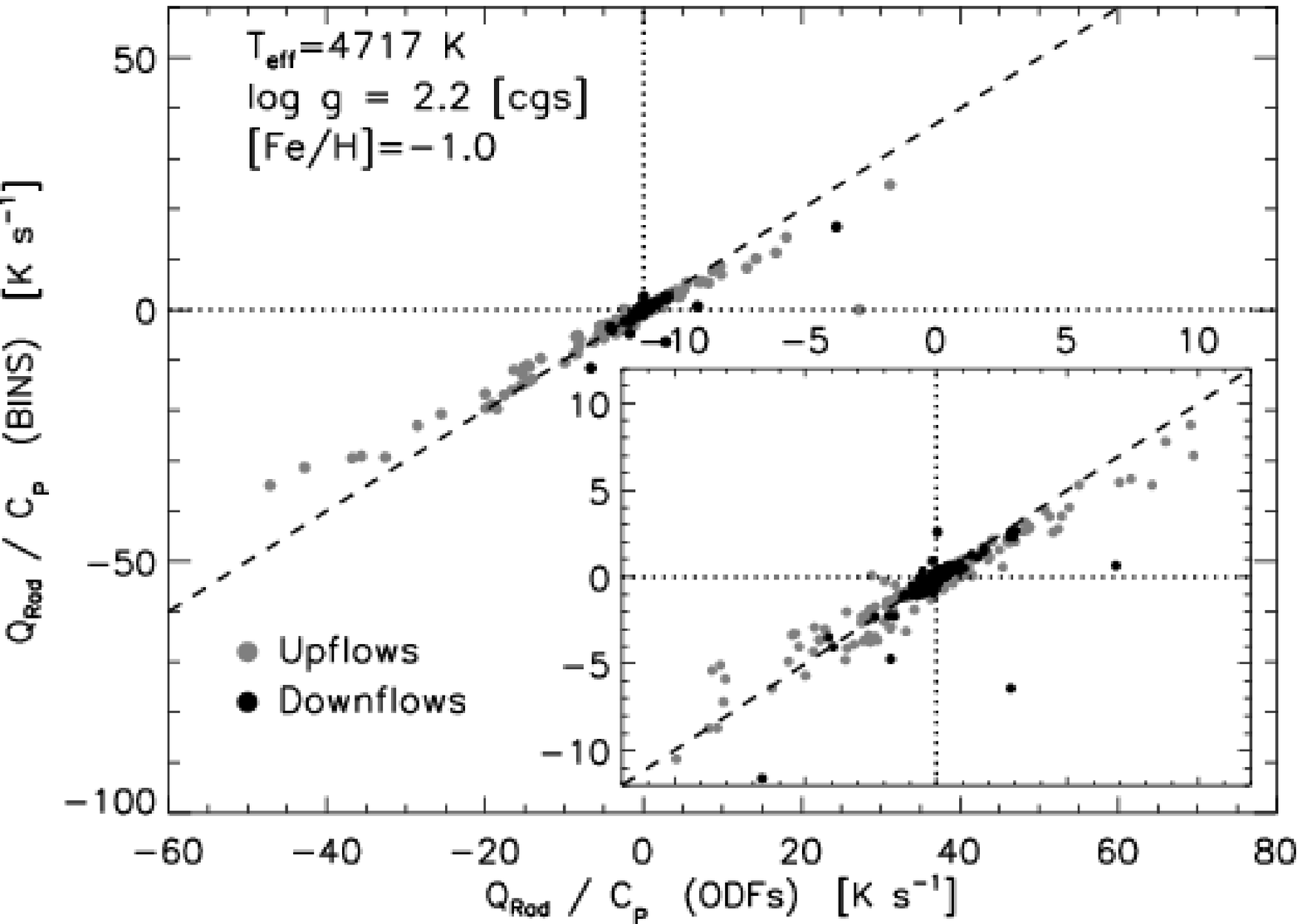} }
   \resizebox{\hsize}{!}{
       \includegraphics{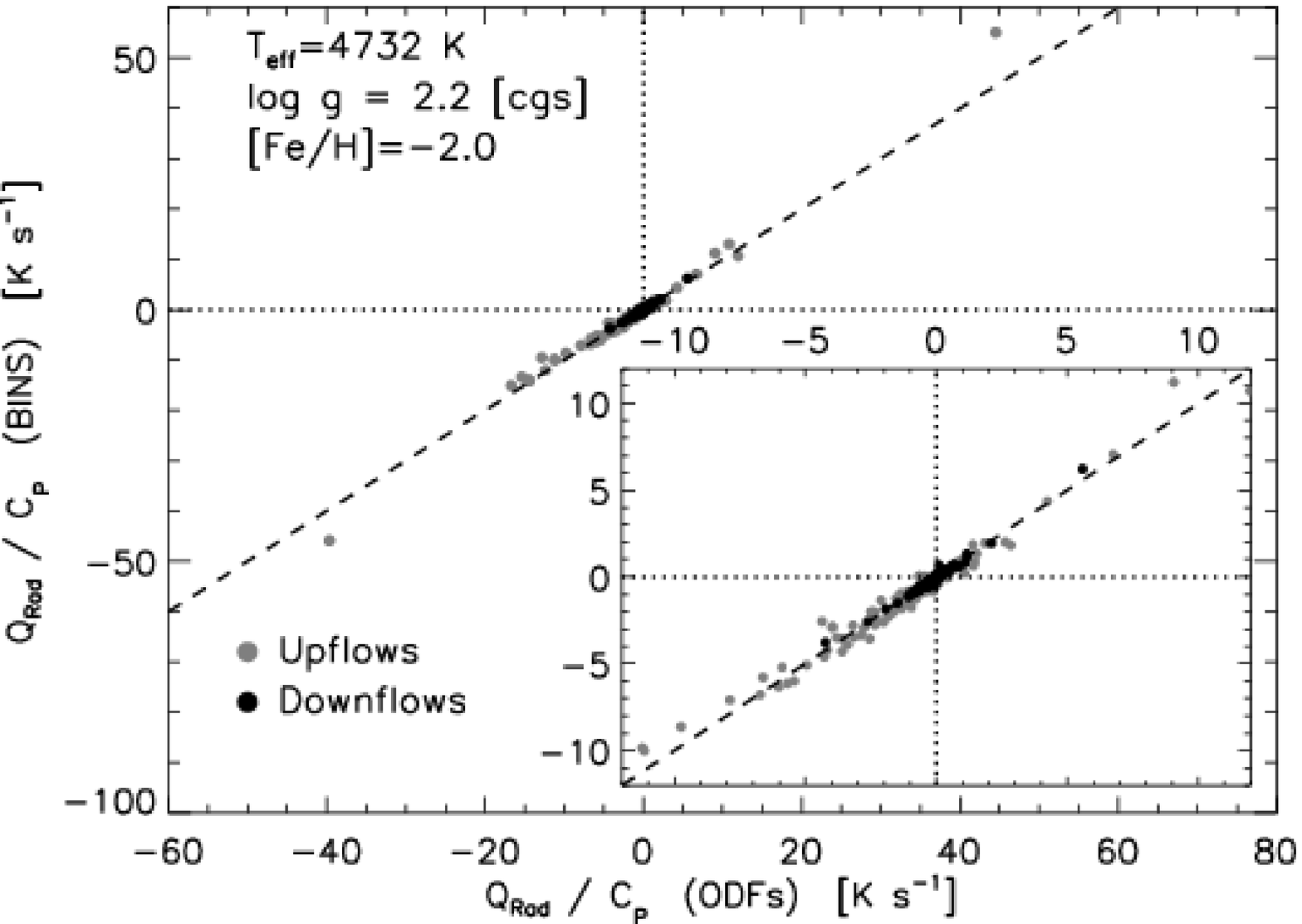}
       \includegraphics{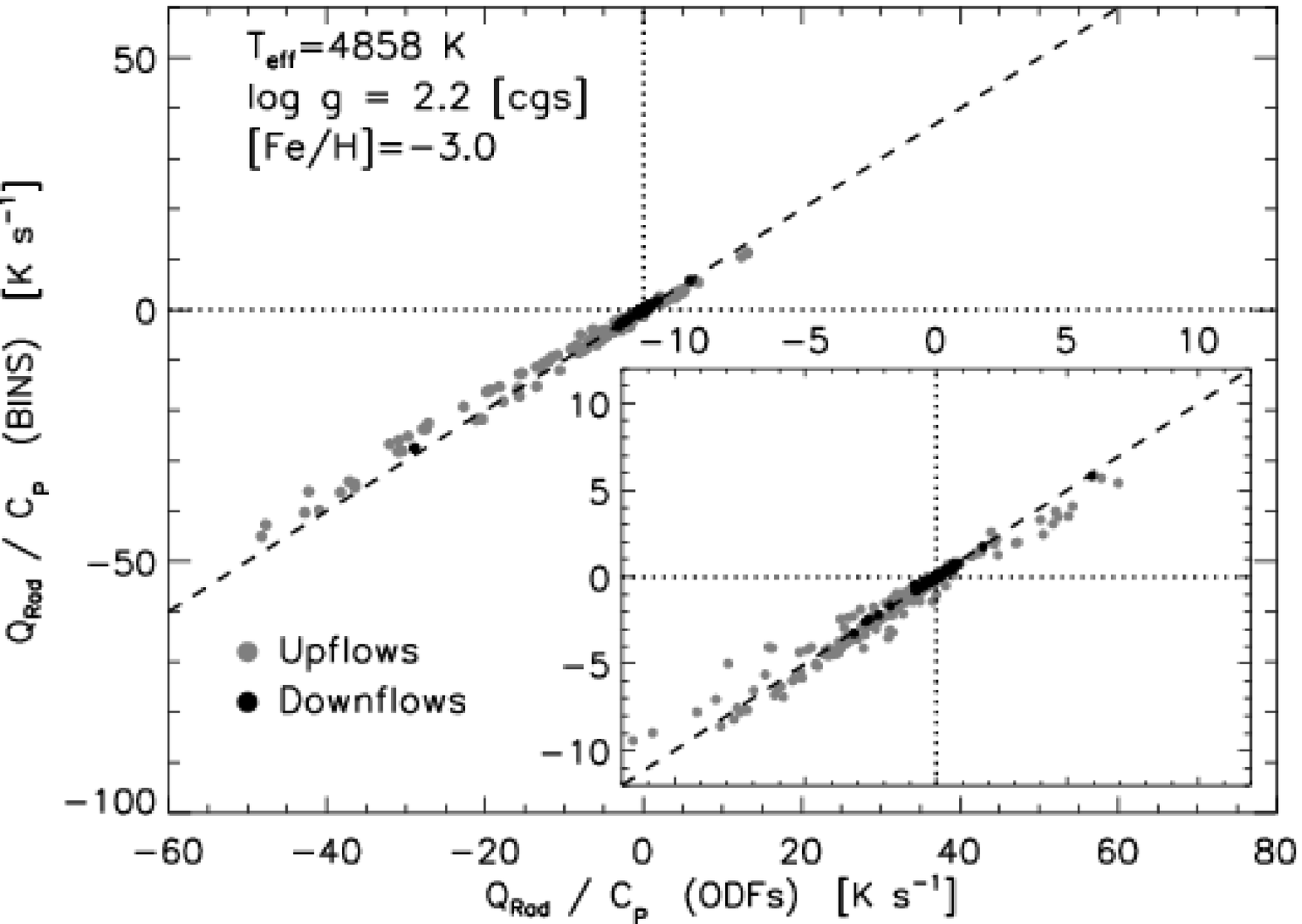} }
   \caption{Comparison between the radiative heating rates 
   computed at all grid-points in four vertical slices of four red giant simulation
   snapshots with different metallicities using the opacity binning scheme
   (``BINS'') and monochromatic radiative transfer with opacity distribution
   functions (``ODFs''). 
   The radiative heating rates per unit mass ($Q_\mathrm{rad}$) are normalized with respect to
   the specific heat (per unit mass) at constant pressure ($C_\mathrm{P}$).
   The small boxes contain magnified views of the plots in the regions of
   low radiative heating rates.}
     \label{fig:models-rqrad}
\end{figure*}

\begin{figure*}
   \centering
   \resizebox{0.9\hsize}{!}{
	\includegraphics{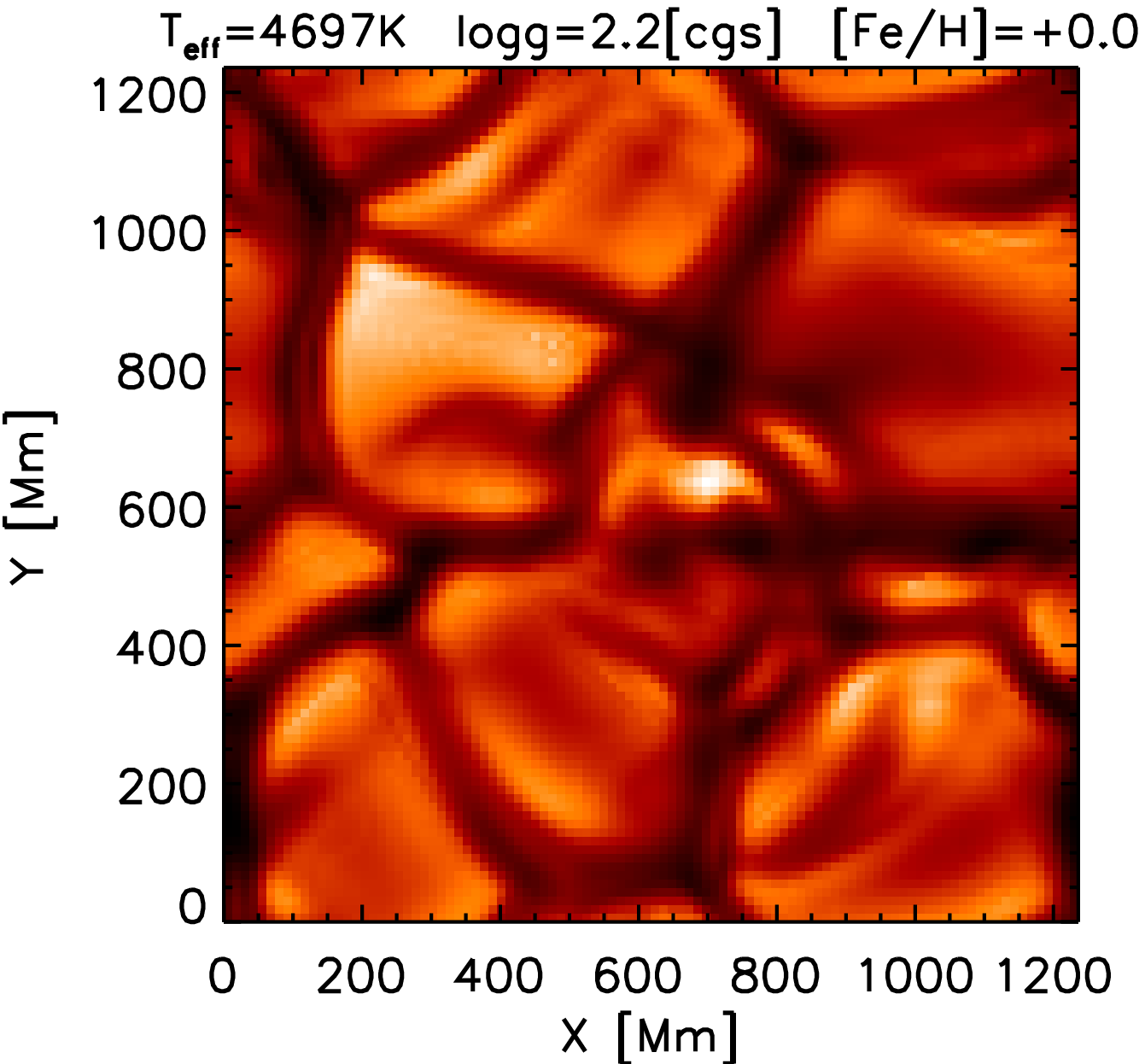}
	\includegraphics{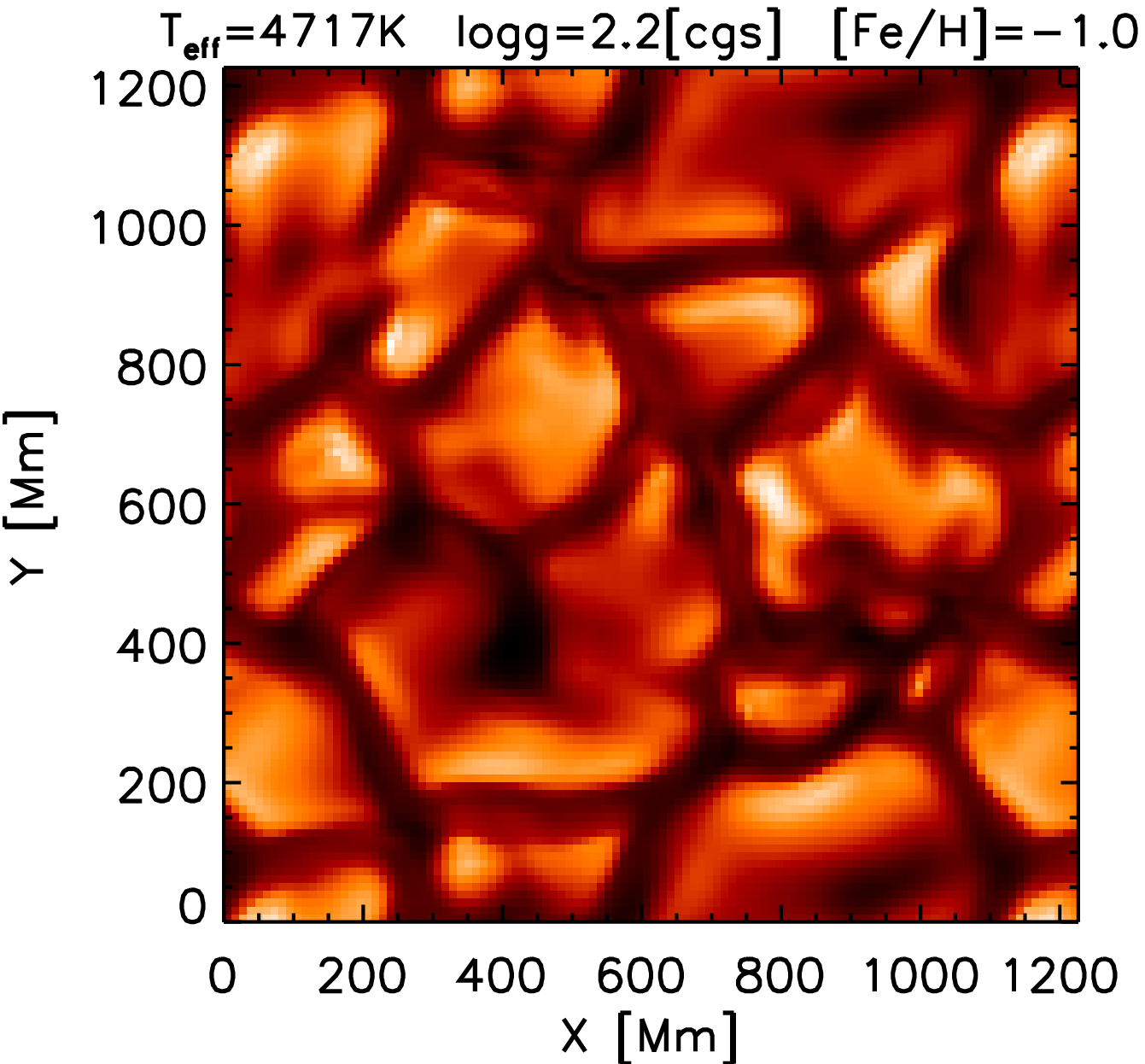} }
   \resizebox{0.9\hsize}{!}{
	\includegraphics{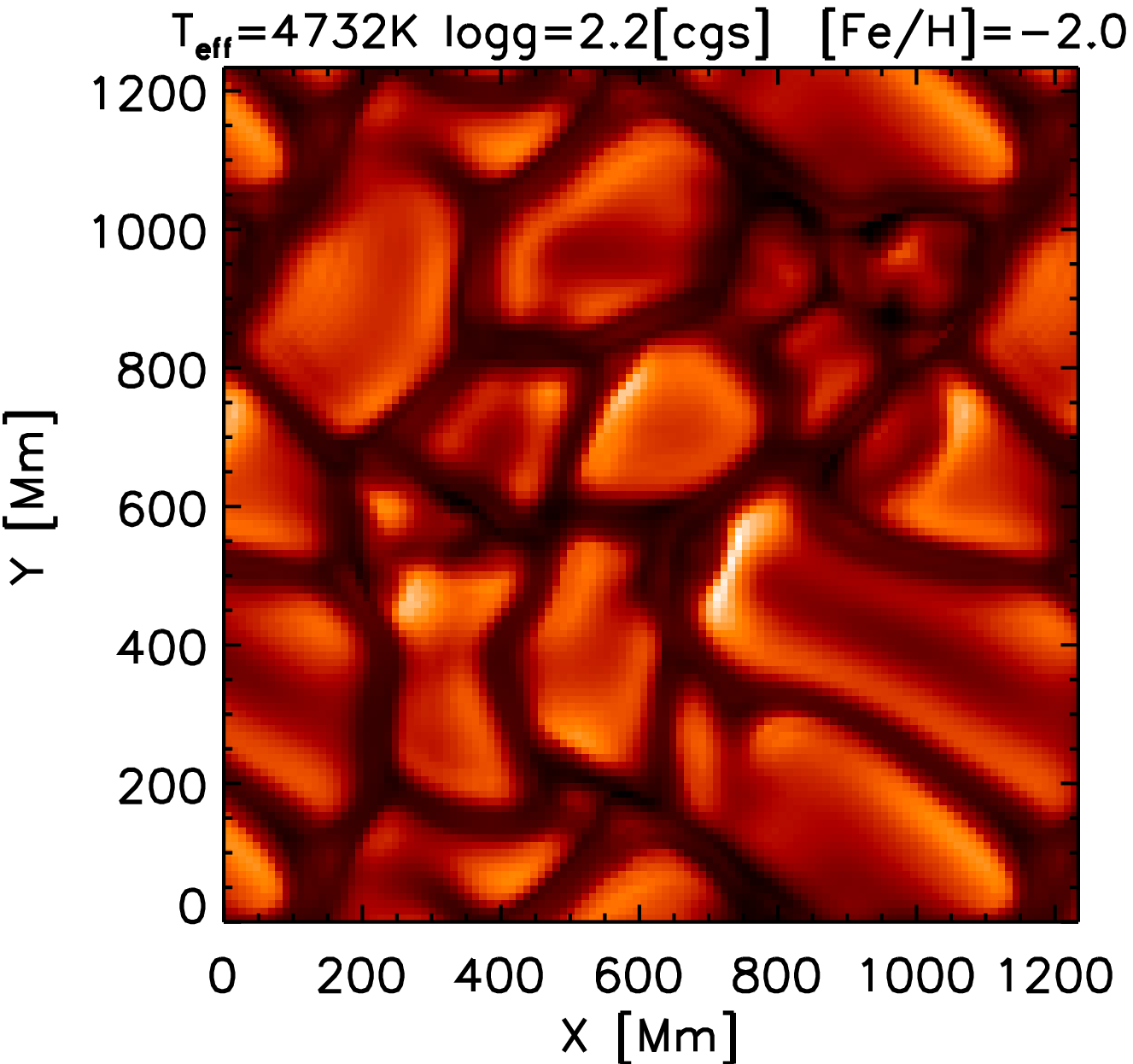}
	\includegraphics{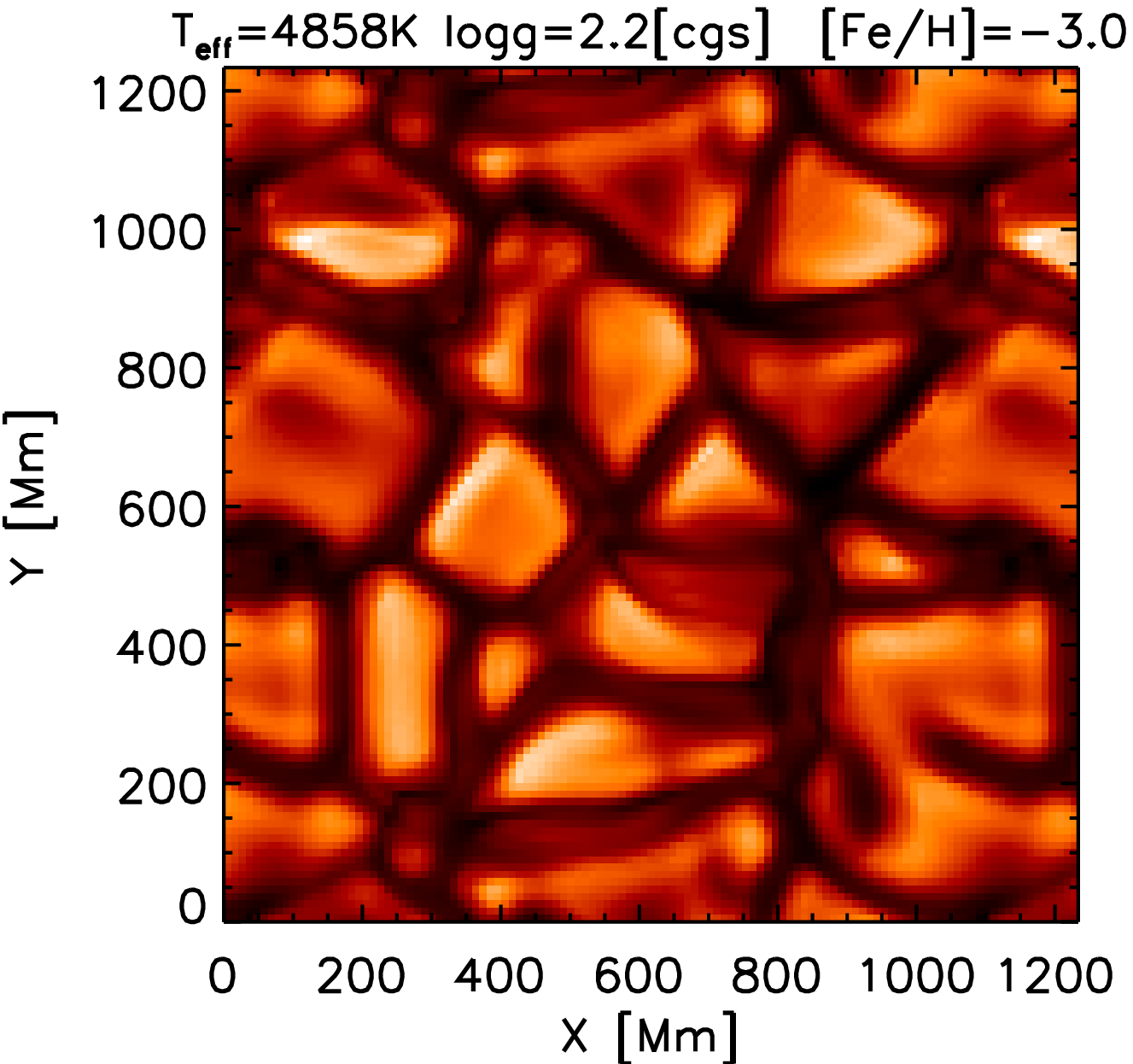} }
   \caption{Spatially resolved emergent intensity in the continuum bin 
   for four snapshots of 3D hydrodynamical simulations of red giants 
   at different metallicities;  
   the characteristic surface granulation pattern is shown. 
   In order to facilitate the comparison of the four cases, 
   the patterns have been partially reproduced periodically
   so that the physical dimensions of the four images are the same as 
   for the $\mathrm{[Fe/H]}=0.0$ simulation.}
  \label{fig:granules}
\end{figure*}

We have also computed classical 1D, LTE, plane-parallel, hydrostatic {\sc marcs}
model atmospheres \citep{gustafsson75,asplund97} with identical stellar
parameters, input data, and chemical compositions as the 3D simulations
to allow for a differential comparison of the two types of models
in terms of spectral line formation and abundance analysis.
For the {\sc marcs} models presented here, we have adopted 
the mixing-length theory (MLT) formulation from \citet{henyey65}, 
with, in particular, the $\alpha_\mathrm{MLT}$ parameter set to $1.5$, 
the structure parameter $y$ to $3/(4\pi^2)\simeq0.076$, 
and the $\nu$ parameter to $0.8$.
We have not considered the effects of turbulent pressure 
when constructing the 1D models.
According to our tests, the inclusion of a turbulent pressure term
$P_\mathrm{turb}=\beta\,\rho\,u^2$, with $u$ the turbulent speed and $\beta=1.5$,
affects the gas pressure-temperature relation but mainly
in those layers below optical depth $\log{\tau_{5000}}\simeq0.4$.
In addition, calculations of synthetic \ion{Fe}{i} and OH line profiles
including and excluding the turbulent pressure term in the 1D models 
indicate that the resulting differences in Fe and O abundances determinations
are typically $0.01$~dex or less.

Contrary to the convection simulations, in the 1D models, 
scattering is correctly treated as such and not as true absorbtion.
The {\sc marcs} models also make use of a slightly different equation of state
\citep[][ and subsequent updates; hereafter ``Uppsala EOS'']{gustafsson73}. 
Compared with the equation of state from \citet{mihalas88}, the
Uppsala EOS predicts for the most part slightly lower values of the gas pressure
at a given gas density and temperature.

The effects of the particular choice of equation of state
on the calculation of the opacity tables are altogether negligible: 
differences in terms of bin opacities computed with the two equations
of state are at most $0.5$~\% at solar metallicity and less than $0.1$~\%  
at $\mathrm{[Fe/H]}=-3$ (for $\log{\tau_{5000}} > -4$).
Our tests also indicate that the differences between 
the 3D$-$1D LTE corrections for Fe and O abundances determined from 
\ion{Fe}{i} and OH lines using the equation of state from \citet{mihalas88}
and the Uppsala EOS are ${\la}0.02$~dex for lines with equivalent 
widths smaller than $80$~m{\AA}. 

Figure~\ref{fig:models} shows the temperature structures
as a function of optical depth resulting 
from the convection simulations at $T_\mathrm{eff}\simeq4750$~K,
compared with the 1D stratification
from the corresponding {\sc marcs} models.
Figure~\ref{fig:models-rho} shows instead the temperature stratifications 
as a function of gas density for both the 3D simulations and 1D 
{\sc marcs} model atmospheres.
At solar metallicities and mild metal-deficiencies ($\mathrm{[Fe/H]}\ga-1$),
the mean temperature-density stratifications at a given optical depth in the upper atmospheric
layers of the hydrodynamical simulations closely resemble
the 1D structures of the corresponding {\sc marcs} models where
radiative equilibrium is enforced.
At lower metallicities ($\mathrm{[Fe/H]}\la-2$), instead,
the temperature stratification in the outer layers of the simulations tends to remain
significantly lower than in 1D model atmospheres.
The temperature in the optically thin layers of the convective 
simulations is for the most part regulated by 
two competing mechanisms: radiative heating caused by reabsorption 
by spectral lines of photons released at deeper layers,
and adiabatic cooling following the expansion of the ascending gas.
Following the interpretation given by \citet{asplund99}
for 3D hydrodynamical models of metal-poor solar-type stars, 
with fewer and weaker lines available at low metallicities, adiabatic
cooling becomes more dominant and the balance between cooling and heating
occurs at lower surface temperatures than in radiative
equilibrium conditions.
At $\mathrm{[Fe/H]}=-3$, the average temperature difference 
between 3D and 1D models in the upper atmospheric 
layers ($\log{\tau_{5000}}\la-3$) is substantial and can amount to $1000$~K or more
at a given optical depth.

The prediction of photospheric temperatures significantly below radiative equilibrium 
values is a crucial result of the convection simulations.
In order to test the accuracy of the opacity binning scheme in this respect,
we computed the radiative heating rates in ``1.5D" approximation 
for all columns in vertical slices of four red giant simulation snapshots 
with varying metallicities using the opacity binning approach 
and monochromatic radiative transfer with opacity distribution functions
(ODFs).  The results are compared in Fig.~\ref{fig:models-rqrad}. 
At very low metallicities ($\mathrm{[Fe/H]}\la-2$) there is an 
excellent correlation between the rates computed with the two approaches, 
suggesting that the opacity binning approximation is accurately 
reproducing the radiative heating and therefore also the temperatures
at the surface of 3D metal-poor models.
At solar metallicity and mild metal-deficiencies ($\mathrm{[Fe/H]}\ga-1$), the correlation
is somewhat poorer, and the results plotted 
in Fig.~\ref{fig:models-rqrad} show that the 
binning scheme generally leads to underestimate both the heating and the
cooling rates compared with monochromatic radiative transfer using ODFs.  
Such weaker response to thermal fluctuations with the opacity binning
approach corresponds to a looser coupling between matter and radiation
which in turn increases the relative importance of convective cooling.
This possibly implies that, at these metallicities, the opacity binning 
scheme is likely to slightly underestimate the temperatures in the upper 
photospheric layers.

Differences between the mean 3D and 1D atmospheric thermal structures,
as well as the presence of temperature and density
inhomogeneities in the 3D hydrodynamical simulations,
can have dramatic effects on the predicted strengths of spectral lines.
The cooler surface layers encountered in the convection
simulations of metal-poor stars are expected to have
a significant impact on temperature sensitive features,
This is the case of, in particular, molecular lines and weak low-excitation 
lines of neutral metals, the line formation regions of which are shifted
outwards in metal-poor hydrodynamical models because of the lower photospheric 
temperatures.
Also, as a consequence of the cooler mean temperature stratifications,
in the upper photospheric layers of metal-poor stars,
the gas and electron pressure resulting 
from the convection simulations tend to remain lower than in
the corresponding 1D hydrostatic models  \citep[see][]{asplund01}.
The lower gas and electron pressures in the metal-poor 3D hydrodynamical
simulations are therefore expected to influence the formation of gravity 
sensitive lines.

Figure~\ref{fig:granules} shows the spatially resolved outgoing 
intensity in the continuum bin for four snapshots of the 
simulations of the $T_\mathrm{eff}{\simeq}4750$~K series.
The granulation pattern is clearly visible and qualitatively resembles
the one observed on the Sun and in other simulations
of late-type stars.
The characteristic size of the granules varies depending on the metallicity
of the simulations. 
The size of granules may be shown to scale approximately with the
pressure scale height $H_P$ \citep[e.g.][]{schwarzschild75}.  
and with the ratio $u_H/u_z$ of horizontal to vertical flow velocities
\citep{stein98}.
At low [Fe/H], the continuous opacity is lower and thus the gas density is
higher at a given optical depth. This means that smaller vertical velocities
are sufficient to sustain the convective flux. With smaller vertical
velocities, horizontal velocities are also expected to be smaller.
In practice, we observe that, in the proximity of the optical surface 
and below it ($\log{\tau_{5000}}\ga-1$), 
the decrease in $u_z$ is overcompensated by a more pronounced decrease 
in $u_H$, so that the product $H_P\,u_H/u_z$ 
becomes in fact systematically larger the higher the metallicity of the simulation,
suggesting that granules must also be bigger.
We also find that the size of the granules increases with the effective temperature of
the simulations.
We intend to return to a more detailed description 
of the physical properties of the convection simulations in a 
future paper.

\subsection{Spectral line formation}
\label{sec:spec-form}
We use the red giant convection simulations as time-dependent
3D hydrodynamical model atmospheres to perform detailed spectral
line formation calculations.
We follow here the same procedure adopted in other recent investigations
of the effects of surface convection and granulation
on solar and stellar spectroscopy \citep{asplund99,asplund00fe1,
asplund00fe2,ags05,asplund00fe3,asplund01,allende01,allende02,nissen02,
collet06}.
From the full red giant simulations, we select representative sequences, 
typically $\sim$100 to $\sim$250 hours long (stellar time), of about 30 snapshots separated 
at regular intervals in time.
Prior to the line formation calculations, we decrease the
horizontal resolution of the simulations from $100{\times}100$ 
down to $50{\times}50$ to ease the computational burden.
We also interpolate the simulations to a finer depth-scale,
increasing the vertical resolution of the layers with 
$\log\tau_{5000}{\la}2.5$ to improve the numerical accuracy.

We compute flux profiles for lines of a number of ions and molecules 
under the assumption of LTE. 
As the main purpose of our study is to isolate and investigate the impact
of 3D models on LTE spectral line formation, we have not performed 
calculations on a large selection of lines ordinarily used in 
abundance analyses.
Instead we consider a sample of ``fictitious'' atomic 
(\ion{Na}{i}, \ion{Mg}{i}, \ion{Ca}{i}, \ion{Fe}{i}, and \ion{Fe}{ii})
and molecular (CH, NH, and OH) lines at selected wavelengths, 
with varying lower-level excitation potentials and line strengths
\citep{steffen02,asplund05,collet06}.
Concerning molecular lines, we restrict our investigation
to a set of low-excitation ($0$~to~$0.5$~eV) features representative of molecular bands
frequently used in abundance analyses of giants 
\citep[e.g.][]{christlieb04,bessell04,cayrel04,spite05}:
CH lines at $4360$~{\AA} belonging to the A$-$X electronic transition band, the OH A$-$X system
at $3150$~{\AA}, and the NH A$-$X lines at $3360$~{\AA}.
Fictitious lines provide a bench-mark for a systematic 
comparison of 1D and 3D LTE spectral line formation 
for various elements and molecules at different metallicities. 
They allow us to analyse the behaviour of spectral lines in 
1D and 3D models uniquely as a function of lower-level excitation potential,
wavelength
and line strength, separating it from other complications such as 
blends or wavelength dependence of continuous opacities.
Finally, in addition to the fictitious spectral lines, 
we also consider some real transitions of particular interest for stellar 
spectroscopy: the \ion{Li}{i} line at $6707.8$~{\AA} and 
the forbidden [\ion{O}{i}] lines at $6300.3$~{\AA} and $6363.7$~{\AA}.

When computing ionization and molecular equilibria and
continuous opacities for the line formation calculations, 
we assume the same chemical compositions as used 
for the construction of the model atmospheres.
As a rule, only the abundances of the trace elements
are varied when calculating line opacities for the 3D cases.
The number densities of line absorbers are then calculated from 
Saha ionization and Boltzmann excitation balances and 
instantaneous molecular equilibrium at the local temperature.
Partition function data for atoms and ions are taken from \citet{irwin81}
and for molecules, together with equilibrium constants, from \citet{sauval84}.

The source function for lines and continuum is approximated with the Planck function
($S_\nu\!=\!B_\nu$) and scattering is treated as true absorption.
We discuss the validity of this approximation for the present analysis
in Sect.~\ref{sec:scattering}.
The radiative transfer equation is solved numerically
along nine directions (two $\mu$-angles, four $\phi$-angles
plus the vertical), after which we perform a disk integration and 
a time average over all selected snapshots.
Various test calculations using a 
larger number of rays (four $\mu$-angles, eight $\phi$-angles plus
the vertical) indicate that, for the scope of the present analysis, 
our procedure is  accurate enough in reproducing
the spatially and temporally averaged line profiles;
the differences between elemental abundances determined with our
procedure and the test cases are typically less than $0.01$~dex.
In terms of derived abundances,
temporal averaging over the selected number of snapshots 
is sufficient to obtain results with the same degree of accuracy,
as verified by test calculations using different numbers of simulation 
snapshots.

To estimate the impact of 3D hydrodynamical models
on stellar spectroscopy, we perform a differential
abundance analysis using a simple curve-of-growth method.
First, we calculate the LTE equivalent widths of the lines of our
sample for a given chemical composition using 1D {\sc marcs} models.
We then evaluate the 3D$-$1D LTE abundance corrections by
varying the abundances in the 3D line formation calculations until the 
equivalent widths reproduce the ones computed in 1D.
Spectral line profiles are calculated for typically 60 to 100 
wavelength points depending on the strength of the lines.
Test calculations performed increasing the spectral resolution
of the line profiles ensure that the accuracy of the differential
3D$-$1D analysis is consistently better than $0.01$~dex
with the adopted number of wavelength points.

Particular care is exerted when dealing with OH and CH lines.
The LTE number densities of the two hydrides in the upper photospheric layers
show a highly non-linear dependence not only on temperature 
but also on the relative abundances of carbon and oxygen. 
Because of the relatively large dissociation energy of the CO molecule (${\sim}11$~eV),
the formation of carbon monoxide can significantly reduce 
the densities of both oxygen and carbon available for 
OH and CH.
Therefore, in order to properly evaluate 3D$-$1D effects
for CH and OH lines, we are forced to take into account
the simultaneous variation of carbon and oxygen abundances
in the 3D line formation calculations.
Here, we determine the 3D$-$1D corrections to \element{C} and \element{O} abundances
self-consistently, by means of an iterative procedure.
The analysis of NH lines, on the contrary, is not affected by such
complications: changes in \element{C} and \element{O} abundances have, in fact,
altogether negligible effects on the strength of NH lines. 
This simplifies the task of determining of 3D$-$1D 
corrections from these lines as variations of the \element{N} abundance 
can be studied independently from the exact value of the \element{C} and \element{O} abundances.
Vice versa, varying the \element{N} abundance in the spectral line formation
calculations causes no appreciable changes on the strengths of CH and OH lines.
 
The same numerical code and input physics are adopted for the line formation 
calculations with both 1D and 3D model atmospheres.
In the 1D cases, we compute spectral line profiles for two different values of 
the micro-turbulence, $\xi=1.5$~km~s$^{-1}$ and $\xi=2.0$~km~s$^{-1}$, 
corresponding to the typical range of values adopted in 
ordinary 1D abundance analyses of red giants with similar 
stellar parameters as the ones adopted for our suites of models.
We emphasize, however, that none of the fudge parameters
needed in the classical 1D analyses (i.e. 
mixing length parameters, micro- and macro-turbulence)
enters the 3D spectral line formation calculations: only the velocity 
fields inherent to the hydrodynamical simulations are
here used to reproduce non-thermal line broadening and asymmetries
associated with convective Doppler shifts. 

For the fictitious lines we implement collisional broadening by hydrogen atoms
according to the classical description by \citet{unsold55} with an
enhancement factor of $2.0$ for the broadening constant.
For the real features analysed here (\ion{Li}{i} and [\ion{O}{i}] lines)
we apply instead the quantum mechanical calculations by \citet{barklem00}.

As we limit our investigation to a differential analysis,
we do not attempt to determine absolute 3D LTE abundances, but
we rather address the question whether systematic effects
are present in classical stellar spectroscopy based on 1D models.
The main advantage of performing a differential study is that the
sensitivity of the results to uncertainties in background continuous
opacities, input physics, absolute line parameters (e.g. lower-level excitation
potentials and $\log{gf}$ values), collisional broadening,
or blends can be minimized.

\begin{figure*}
   \centering
   \resizebox{\hsize}{!}{ \includegraphics{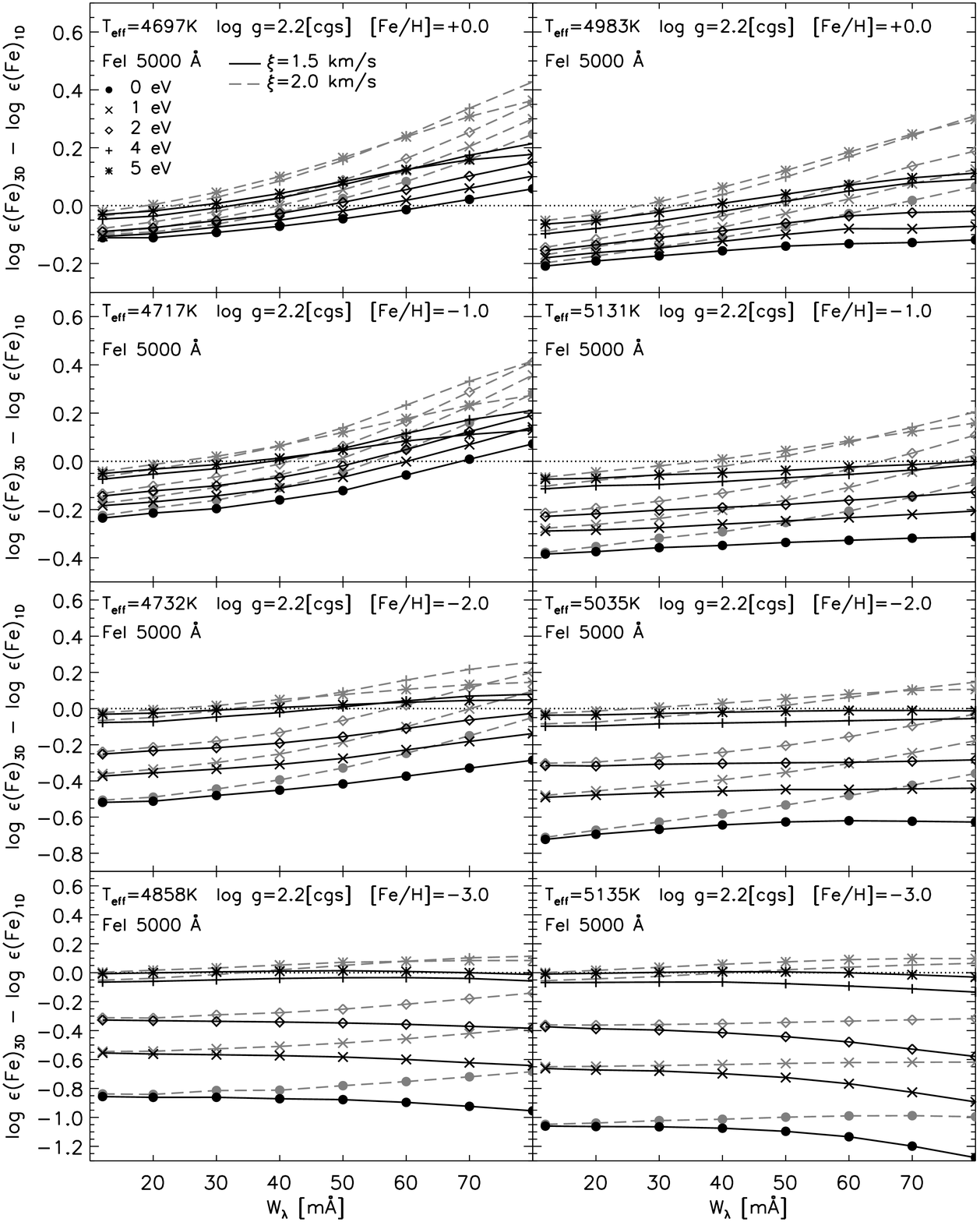} }
   \caption{3D$-$1D LTE corrections to \element{Fe} abundances derived
	from \ion{Fe}{i} fictitious lines at $\lambda=5000$~{\AA} 
	as a function of equivalent width $W_\lambda$.}
   \label{fig:fei5000}
\end{figure*}

\begin{figure*}
   \centering
   \resizebox{\hsize}{!}{\includegraphics{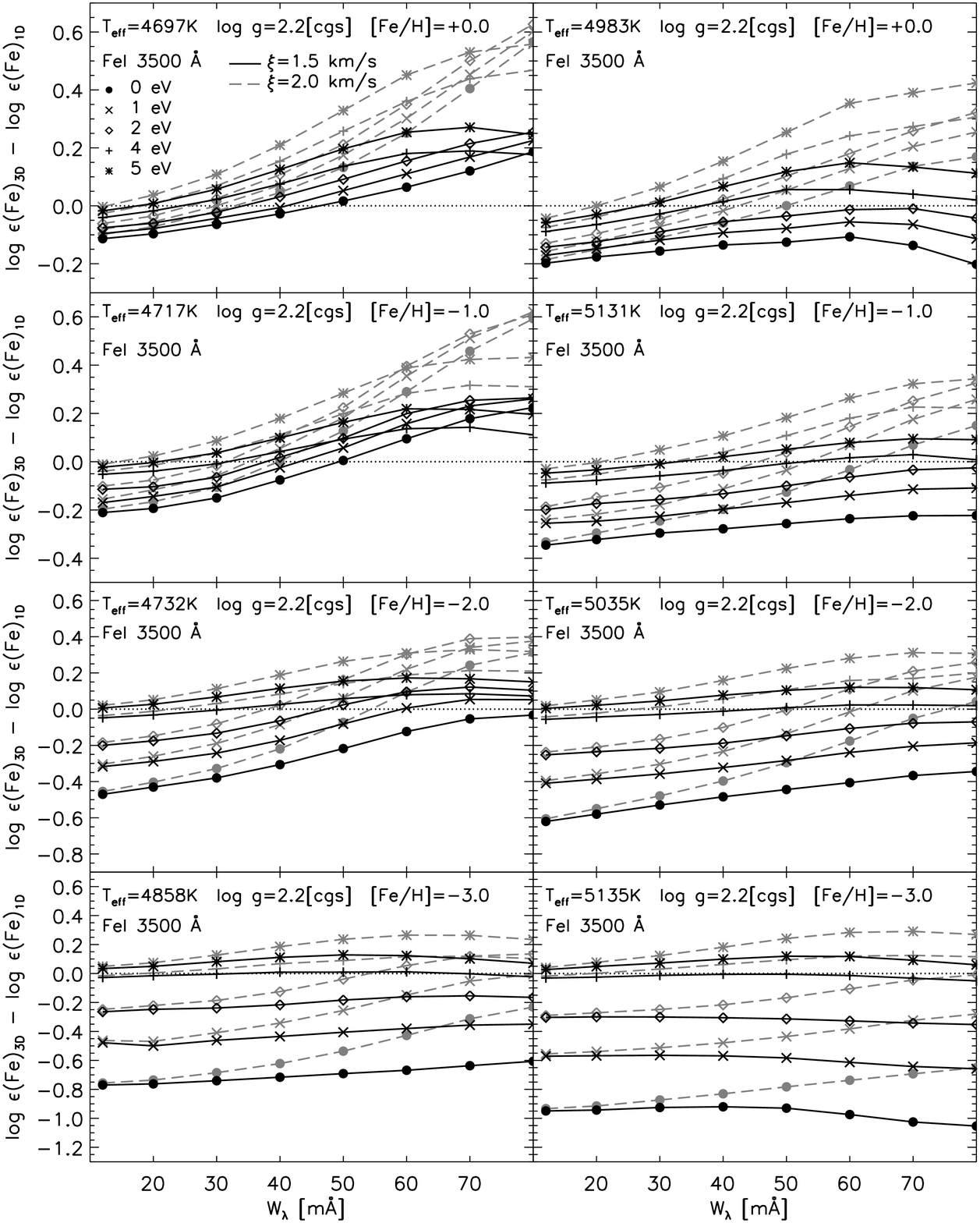}}
   \caption{3D$-$1D LTE corrections to \element{Fe} abundances derived
	from \ion{Fe}{i} fictitious lines at $\lambda=3500$~{\AA} 
	as a function of equivalent width $W_\lambda$.}
   \label{fig:fei3500}
\end{figure*}

\begin{figure*}
   \centering
   \resizebox{\hsize}{!}{ \includegraphics{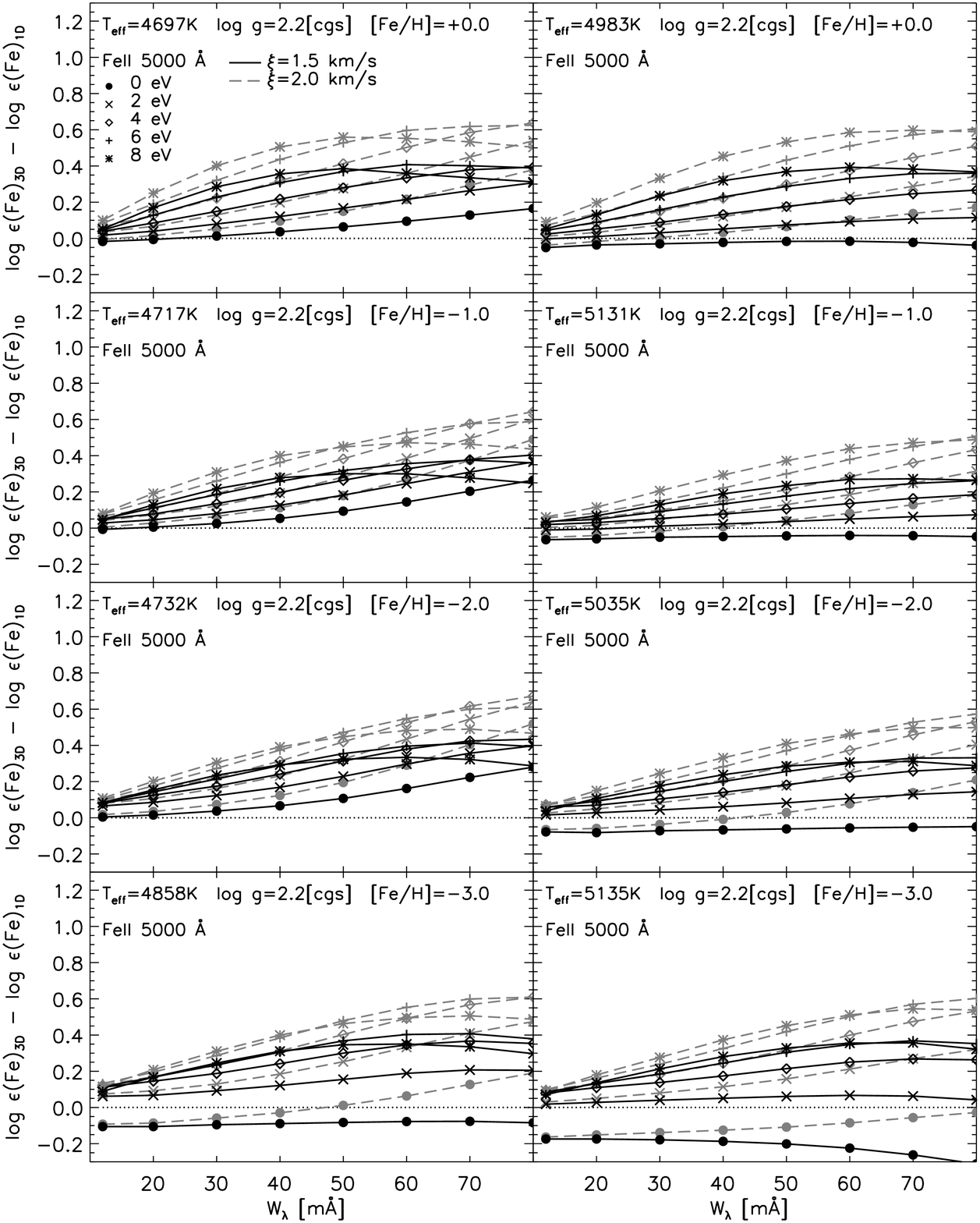} }
   \caption{3D$-$1D LTE corrections to \element{Fe} abundances derived
	from \ion{Fe}{ii} fictitious lines at $\lambda=5000$~{\AA}
	as a function of equivalent width $W_\lambda$.}
   \label{fig:feii5000}
\end{figure*}

\begin{figure*}
   \centering
   \resizebox{\hsize}{!}{\includegraphics{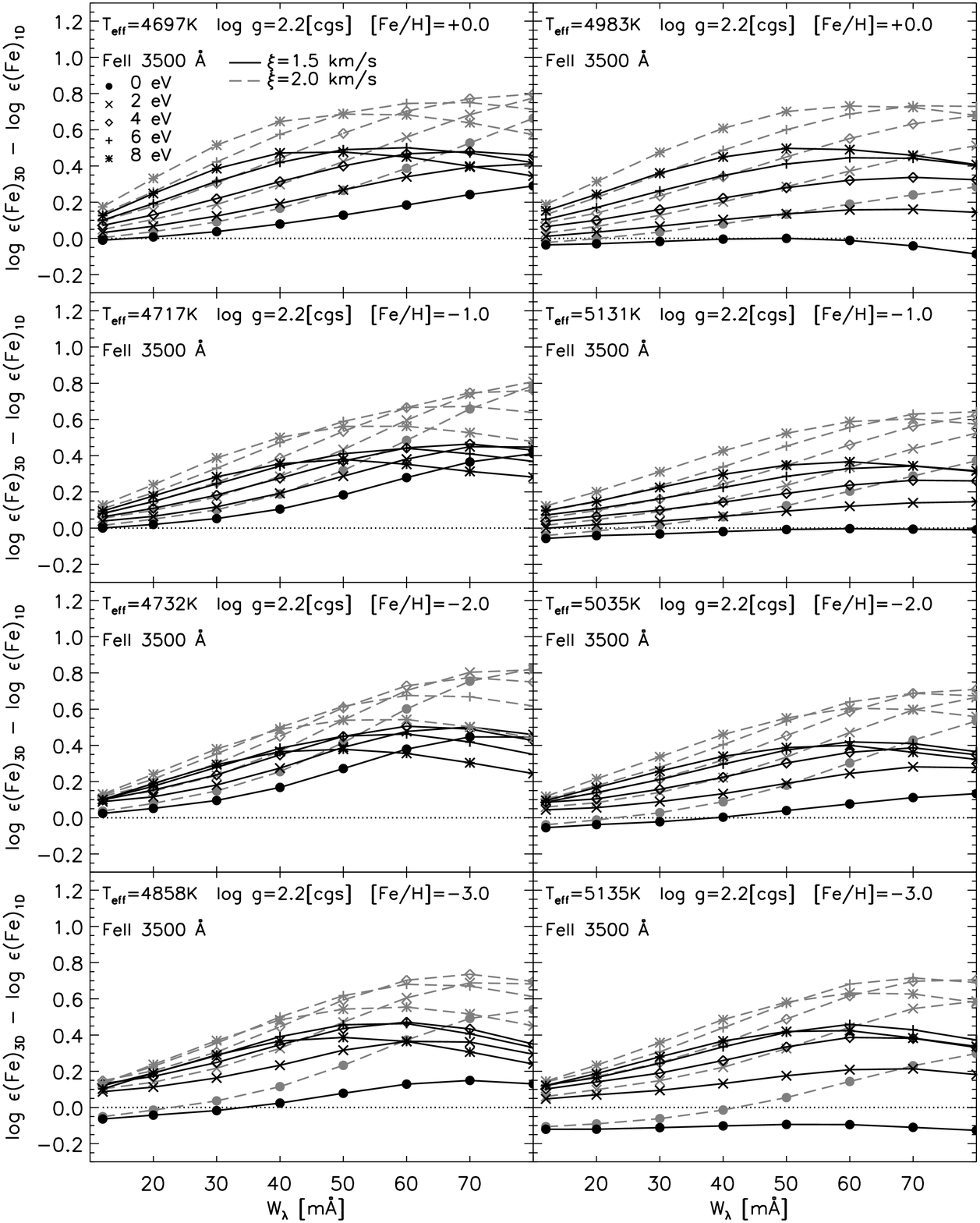}}
   \caption{3D$-$1D LTE corrections to \element{Fe} abundances derived
	from \ion{Fe}{ii} fictitious lines at $\lambda=3500$~{\AA}
	as a function of equivalent width $W_\lambda$.}
   \label{fig:feii3500}
\end{figure*}

\begin{figure*}
   \centering
   \resizebox{\hsize}{!}{ \includegraphics{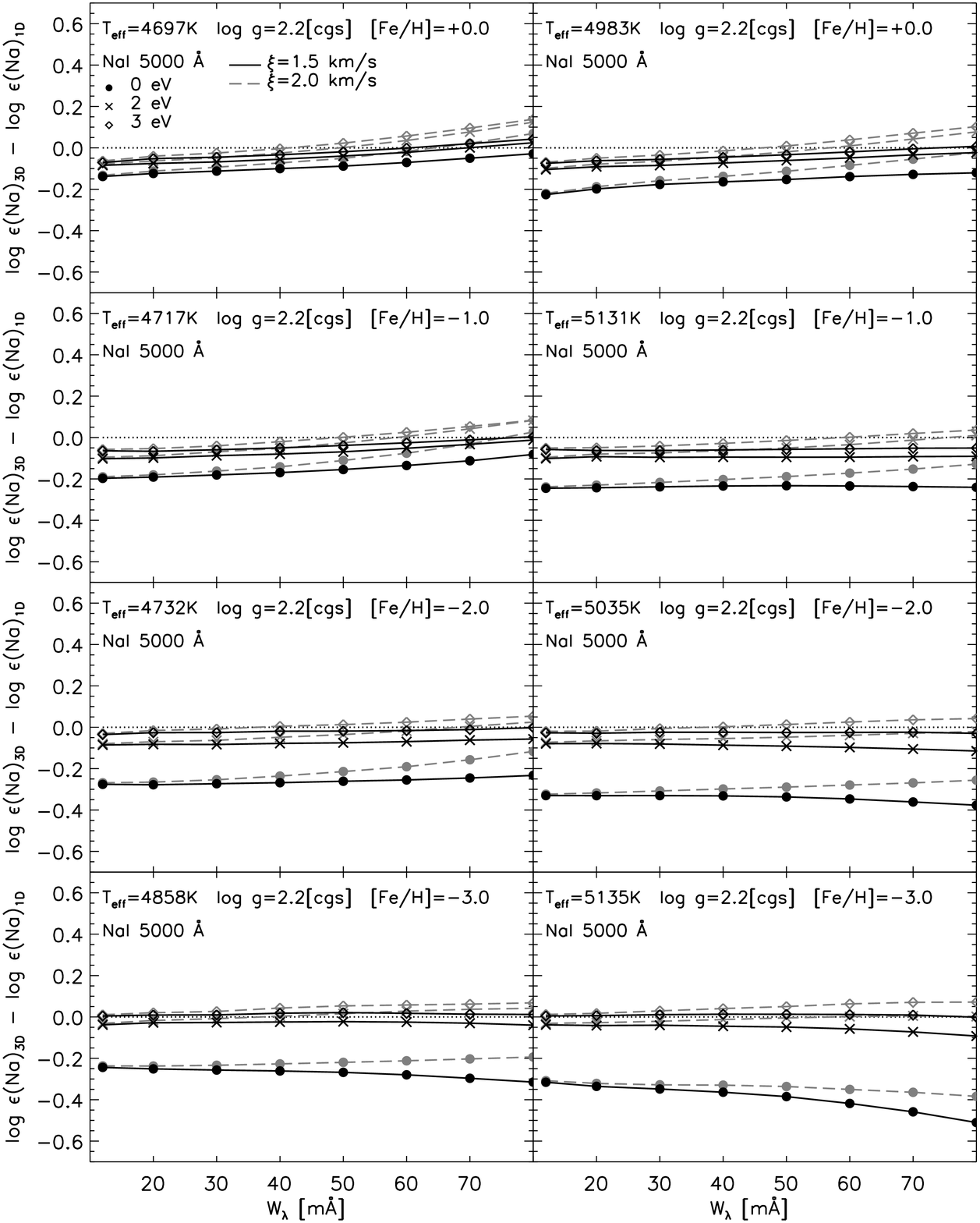} }
   \caption{3D$-$1D LTE corrections to \element{Na} abundances derived
	from \ion{Na}{i} fictitious lines at $\lambda=5000$~{\AA}
	as a function of equivalent width $W_\lambda$.}
   \label{fig:nai5000}
\end{figure*}

\begin{figure*}
   \centering
   \resizebox{\hsize}{!}{ \includegraphics{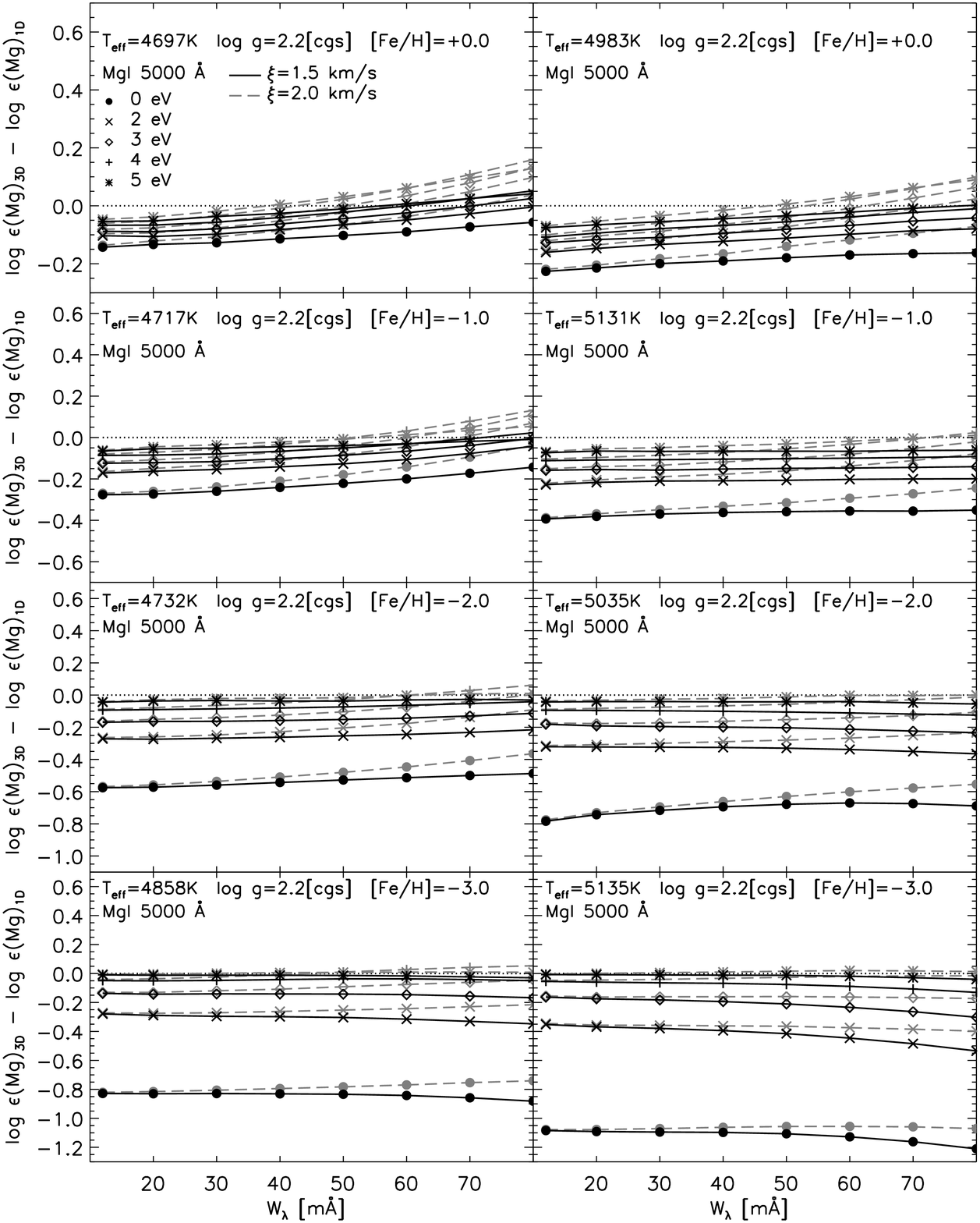} }
   \caption{3D$-$1D LTE corrections to \element{Mg} abundances derived
	from \ion{Mg}{i} fictitious lines at $\lambda=5000$~{\AA}
	as a function of equivalent width $W_\lambda$.}
   \label{fig:mgi5000}
\end{figure*}

\begin{figure*}
   \centering
   \resizebox{\hsize}{!}{ \includegraphics{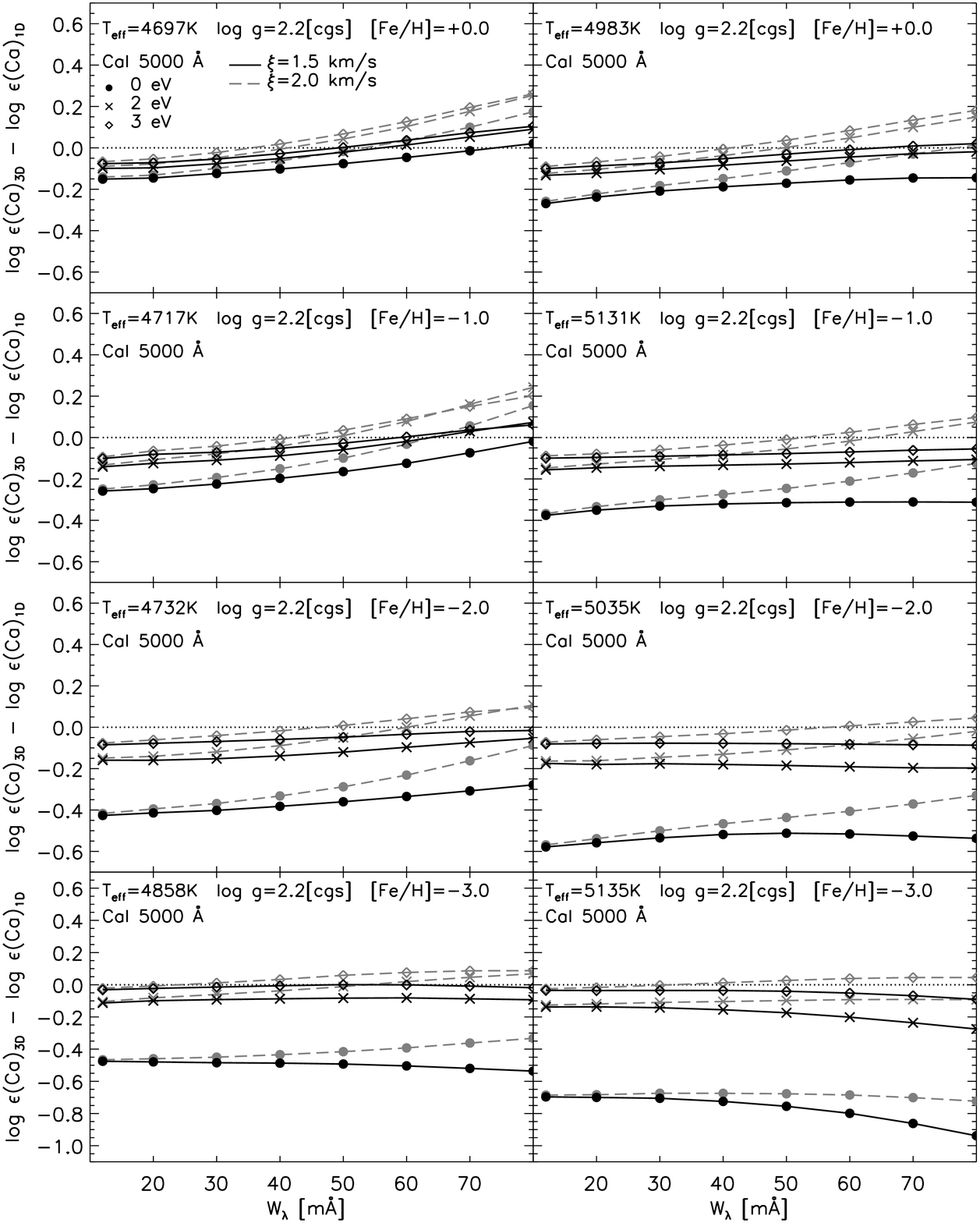} }
   \caption{3D$-$1D LTE corrections to \element{Ca} abundances derived
	from \ion{Ca}{i} fictitious lines at $\lambda=5000$~{\AA}
	as a function of equivalent width $W_\lambda$.}
   \label{fig:cai5000}
\end{figure*}

\section{Results}
\subsection{\ion{Fe}{i} and \ion{Fe}{ii} lines}
\label{sec:results-fe}
Figure~\ref{fig:fei5000} shows the 3D$-$1D LTE corrections to the 
\element{Fe} abundance as derived, for the two series of model atmospheres,
from \ion{Fe}{i} lines  at $\lambda=5000${\AA} with varying equivalent widths $W_\lambda$.
At a given wavelength, the magnitudes of the corrections and their trends 
as a function of line strength depend on several parameters such as 
lower-level line excitation potential, metallicity, and
effective temperature of the models.

At a given \element{Fe} abundance, weak low-excitation \ion{Fe}{i} lines
appear \emph{stronger} in the framework of 3D models than they do in 1D,
resulting in negative 3D$-$1D LTE abundance corrections.
The effects are more pronounced at lower metallicities where the differences 
between the 3D and 1D \element{Fe} abundances can reach $\sim-1.0$~dex.
The 3D$-$1D abundance corrections become progressively smaller for 
higher-excitation lines, vanishing for line excitation potentials
of approximately $5$~eV. 
A similar result characterizing granulation abundance corrections
for the Sun is reported by \citet{steffen02}. 
In their spectroscopic analysis of the red giant Pollux ($\beta$ Gem), 
\citet{ruland80} find that abundances derived for low-excitation lines are typically
lower than for high-excitation lines and interpret the result
as an effect of departures from LTE.
Applying our 3D$-$1D corrections, the trend of the derived abundances as a function
of excitation potential would become even steeper, suggesting that the non-LTE
effects might be even larger for the 3D case. 
Possible departures from LTE of \element{Fe} are discussed further in Sect.~\ref{sec:nlte}

The overall behaviour of the 3D$-$1D abundance corrections as a function
line strength varies significantly depending on the metallicity of the models.
At solar and moderately low metallicities ($\mathrm{[Fe/H]}\ga-1$),
the differences between 3D and 1D abundances become progressively 
less negative with increasing line strength
and may eventually revert sign, growing positive for sufficiently strong lines 
and relatively high excitation potentials.
It is also worthwhile observing that the abundance corrections for strong lines
can reach a maximum and then turn-over as the line excitation potential increases.
At lower metallicities, the 3D$-$1D abundance corrections generally show a shallower trend
with equivalent width.
In particular, at $\mathrm{[Fe/H]}=-3$, the corrections
can also grow more negative for stronger lines.
The values of the 3D$-$1D abundance corrections for relatively strong lines 
($W_\lambda\ga50$~{m\AA}) also depend on the choice of micro-turbulence,
which controls the saturation level of the lines in the 1D calculations.
At a given abundance in 1D the larger the micro-turbulence,
the stronger the lines and in turn the less negative (or more positive)
the differences between 3D and 1D abundances become.

Figure~\ref{fig:fei3500} shows analogous 3D$-$1D \element{Fe} abundance corrections 
computed for \ion{Fe}{i} lines at $3500$~{\AA}.
The results are fairly similar to the ones for \ion{Fe}{i} lines at
$5000$~{\AA}, the trends of the corrections following
qualitatively the same behaviour as a function of line strength.
 The exact values of the corrections can vary slightly for the
two wavelengths due to the different steepness of the source function
with depth and different sensitivity of the continuous opacity
at $3500$~{\AA} and $5000$~{\AA} to temperature and electron density.

In general, identifying the leading reason for a particular behaviour
of the 3D$-$1D corrections as a function of the line or the model parameters
is not a simple task, due to the complexity and non-linearities inherent
to line formation, in particular in 3D hydrodynamical models.
In practice, spectral lines form over a relatively large
range of optical depths; in the framework of 3D models line
formation is sensitive not only to the mean thermal stratification
but also to inhomogeneities in the temperature, density, and velocity fields.
The large abundance corrections for weak lines in the very metal-poor cases
can be qualitatively understood by comparing the 3D and 1D model structures.
The mean temperature stratifications at the surface of 3D hydrodynamical
simulations of very metal-poor giants are significantly cooler than in 1D models
computed for the same stellar parameters (Sect.~\ref{sec:convec-sim}).
This implies that in LTE the fraction of neutral to ionized iron is overall
enhanced in the upper layers of 3D metal-poor models compared with the 1D case.
The cooler temperature structure of 3D models also 
contributes to reducing 
the electron pressure and, consequently, lowering the density of \element{H}$^-$ ions,
thus decreasing the opacity in the continuum-forming layers.
Therefore, at low metallicities, the combined effect of increased line opacity and
decreased continuous opacities tends to make weak low-excitation
\ion{Fe}{i} lines stronger in 3D than in 1D; a lower \element{Fe} abundance 
is then required in metal-poor 3D models to reproduce the 1D equivalent widths.

Weak high-excitation lines form in deeper photospheric layers and
are less sensitive to the temperature structure at the surface of
the models. 
At all metallicities, the resultant 3D$-$1D abundance corrections for high-excitation
lines are smaller than for the low-excitation features.
Stronger lines form higher up in the atmosphere compared with weak lines 
when the other line properties remain the same.
While at very low metallicities \ion{Fe}{ii} is clearly the dominant
\element{Fe} ionization stage nearly everywhere in the 1D models,
in 3D the \ion{Fe}{i} fraction is substantial at all depths
contributing to the emergent line profiles.
Therefore in very metal-poor models the 3D$-$1D \element{Fe} abundance corrections
typically remain negative for strong \ion{Fe}{i} lines.
As a test, we compute curves-of-growth of \ion{Fe}{i} lines using
the horizontally averaged 3D structure to quantitatively study
the dependence of line strengths on the lower temperatures of
the mean stratification.
The behaviour of the  3D$-$1D corrections for  \ion{Fe}{i} lines
can be qualitatively reproduced by comparing the curves-of-growth 
for the 1D {\sc marcs} and mean 3D stratifications. 
The test confirms that the lower temperatures encountered 
in the 3D hydrodynamical metal-poor model atmospheres are the 
main factor determining the large and negative  3D$-$1D corrections.
However, we emphasize that, from a quantitative point of view,
line formation calculations relying on the mean  3D stratification 
cannot accurately reproduce the results of the calculations
based on the actual 3D model atmospheres.
In fact, abundances derived from weak low excitation  \ion{Fe}{i} lines
using the full 3D and the mean 3D stratifications can differ by
up to about $0.3$~dex.
This is an indication that the temperature and velocity inhomogeneities 
of the 3D structure should not be neglected.

At metallicities closer to solar, the trends of 3D$-$1D \element{Fe} abundance corrections
with line strength cannot be ascribed simply to differences between
1D and mean 3D stratifications, which are very similar
in the line forming layers.
Curves-of-growth computed for \ion{Fe}{i} lines
using either the 1D {\sc marcs} models or the mean 3D stratifications 
(and the same value of micro-turbulence) closely resemble each other. 
The differences in \element{Fe} abundance determinations using
the 1D and mean 3D stratifications amount in fact to less than $0.04$~dex
for \ion{Fe}{i} lines within the range of line strengths
considered here.
Hence, the actual trends of the 3D$-$1D corrections
at metallicities near solar must be attributed mainly to 3D temperature inhomogeneities
and possibly velocity fields.
We also note that the trend of the 3D$-$1D abundance corrections,
increasing with equivalent width for $[\mathrm{Fe/H}]{\ga}-1$, 
can be flattened by appropriately choosing a smaller
value for the micro-turbulence than the ones considered here ($\xi<1.5$~km~s$^{-1}$).
However, typical micro-turbulences derived from 
spectroscopic 1D analyses of red giants are found indeed to be in the range 
between $1.5$~km~s$^{-1}$ and $2.0$~km~s$^{-1}$. 
At metallicities $[\mathrm{Fe/H}]{\ga}-1$, 
the present 3D$-$1D \element{Fe} abundance corrections,
would therefore introduce a trend in the derived abundances as a function
of equivalent width.
This could be an indication that convective motions of the gas are not
fully resolved in the present hydrodynamical simulations and, consequently, 
that non-thermal Doppler broadening is also underestimated.
Further investigation is however necessary in order to identify other
possible causes for these trends.

We observe that
the 3D$-$1D \element{Fe} abundance corrections for lines at a given 
equivalent width are typically shifted toward more negative values 
for the higher effective temperature models.
The equivalent width of a (weak) spectral line is roughly proportional to 
the ratio $\ell_\nu/\kappa_\nu$ of line to continuous opacity 
at line formation depth \citep[see also][~p.~277]{gray92}.
In the case of (weak) low-excitation \ion{Fe}{i} lines, for a given 
\element{Fe} abundance, the ratio 
$(\ell_\nu/\kappa_\nu)_\mathrm{3D}/(\ell_\nu/\kappa_\nu)_\mathrm{1D}$
turns out to be larger the higher the effective temperature of
the models at all metallicities, indicating that 3D$-$1D \element{Fe} 
abundance corrections are also expected to be more pronounced 
(more negative).

The 3D$-$1D \element{Fe} abundance corrections for \ion{Fe}{ii} lines
at $5000$~{\AA} and $3500$~{\AA} are presented in Fig.~\ref{fig:feii5000} and \ref{fig:feii3500}.
The overall trends of the corrections for these lines are similar at all metallicities;
differences between abundances derived with 3D and 1D models are typically
positive and generally increase with increasing line strength and excitation potential. 
Low-excitation lines can, however, show negative corrections at metallicities
below $\mathrm{[Fe/H]}\simeq-2$.
Similarly as for \ion{Fe}{i} lines also, the corrections 
\ion{Fe}{ii} lines at a given equivalent width can reach a maximum and revert
their trend for sufficiently high excitation potentials. 
This behaviour is also present in 3D$-$1D abundance corrections for
lines of neutral and ionized metals in solar-type stars
\citep{asplund05}, although evident already for somewhat weaker lines.

\subsection{\ion{Na}{i}, \ion{Mg}{i}, and \ion{Ca}{i}}
\label{sec:namgca}

Figures~\ref{fig:nai5000}, \ref{fig:mgi5000}, and \ref{fig:cai5000} show
the 3D$-$1D LTE corrections to the sodium, magnesium, and calcium  
abundances as derived from \ion{Na}{i}, \ion{Mg}{i}, and \ion{Ca}{i} lines at $5000$~{\AA}.
The behaviour of the abundance corrections as a function 
of line strength for these lines is  similar to the one predicted 
for \ion{Fe}{i} lines.
In LTE, the overall trends of the corrections are governed by the 
differences between the 1D and mean 3D temperature stratifications and
by the presence of temperature inhomogeneities and velocity fields in the
3D models.
The actual magnitude of the 3D$-$1D effects, however, largely depends on 
the ionization potential of the species under consideration.
The 3D$-$1D abundance corrections for \ion{Na}{i} and \ion{Ca}{i}, for instance, are
considerably smaller than the ones derived from \ion{Fe}{i} lines.
In fact, the relatively low ionization potentials of \ion{Na}{i} and \ion{Ca}{i}
imply that the two metals are more easily ionized compared with \ion{Fe}{i},
both in 1D and 3D models. This significantly reduces the 3D$-$1D effects on line strengths 
due to temperature inhomogeneities or differences in the thermal stratifications.
Magnesium on the contrary has a fairly high first ionization potential, comparable
to that of iron: hence, the computed 3D$-$1D abundance corrections for \ion{Mg}{i}
lines are typically also as large as the ones derived from  \ion{Fe}{i} lines.

\subsection{\ion{Li}{i}}

\begin{table}
\caption{Differential 3D and 1D \element{Li} abundances derived 
from the \ion{Li}{i}~$6707.8$~{\AA} line.
The computed equivalent widths $W_\lambda$ of the \ion{Li}{i} line 
are also reported.
A $\log{gf}$ value of $0.174$ is assumed \citet{smith98} in
the line formation calculations.
}
\label{tab:li-corr}
\centering
\begin{tabular}{cc c c c cc}
\hline\hline 
$\langle T_\mathrm{eff}\rangle$&
$[\mathrm{Fe/H}]$ & 
& 
$\log{\epsilon(\element{Li})}_\mathrm{1D}$ &
$\log{\epsilon(\element{Li})}_\mathrm{3D}$ &
&
$W_\lambda$ \\
   
$[\mathrm{K}]$	& & & &  && $[\mbox{m{\AA}}]$ \\ 
\hline
$4697$ &  $+0.0$   &  & 1.0 &  0.90 & &  37.7\\
$4717$ &  $-1.0$   &  & 1.0 &  0.77 & &  34.7\\
$4732$ &  $-2.0$   &  & 1.0 &  0.62 & &  30.1\\
$4858$ &  $-3.0$   &  & 1.0 &  0.55 & &  22.0\\
\noalign{\smallskip}		    
$4983$ &  $+0.0$   &  & 1.0 &  0.78 & &  17.5\\
$5131$ &  $-1.0$   &  & 1.0 &  0.66 & &  11.7\\
$5035$ &  $-2.0$   &  & 1.0 &  0.56 & &  14.7\\
$5128$ &  $-3.0$   &  & 1.0 &  0.50 & &  12.7\\
\hline 
\end{tabular}
\end{table}

\begin{figure}
   \centering
   \includegraphics[width=0.5\textwidth]{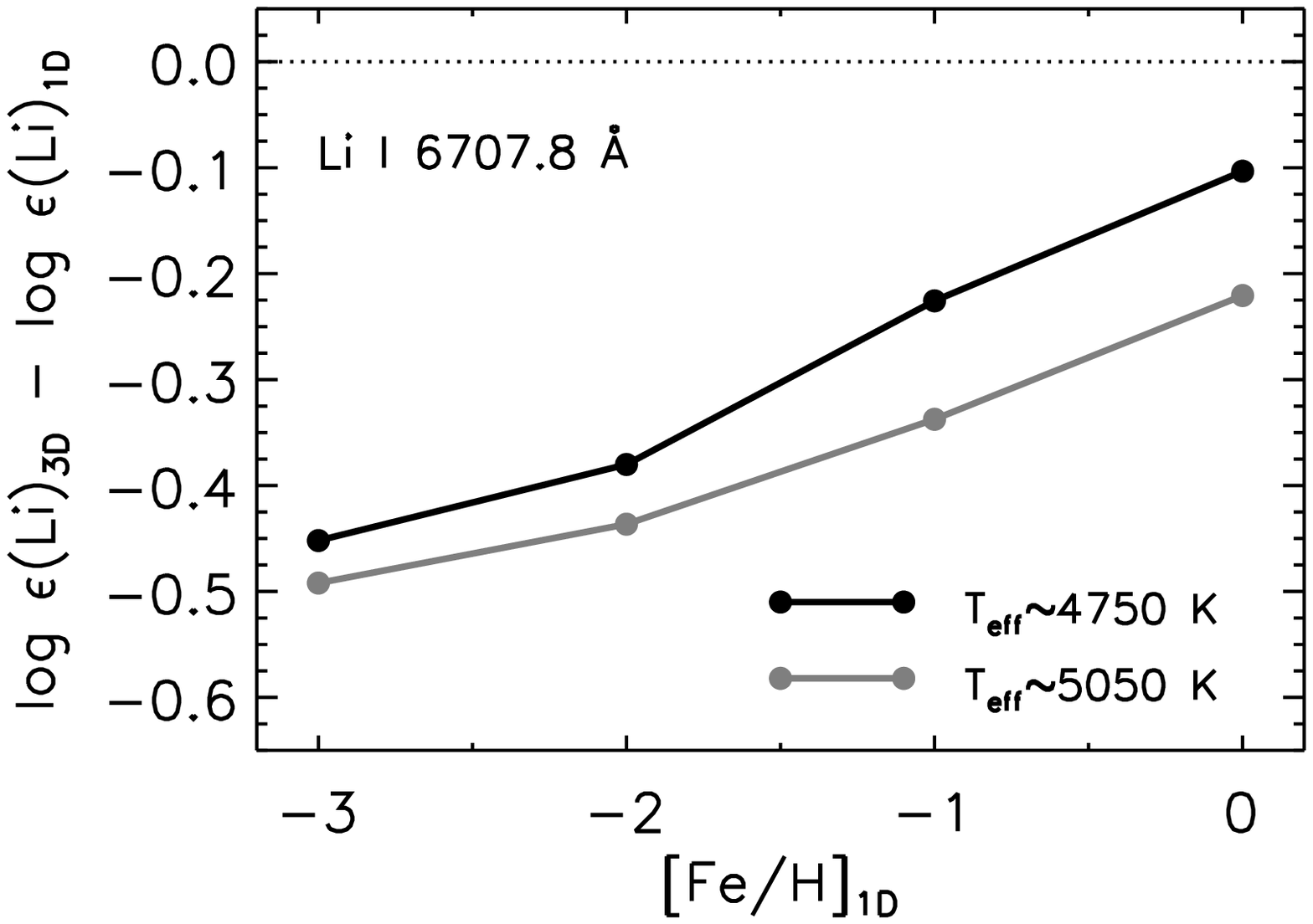} 
   \caption{3D$-$1D LTE corrections to \element{Li} abundances derived
	from the \ion{Li}{i} line at $\lambda=6707.8$~{\AA} 
	as a function of metallicity $[\mathrm{Fe/H}]$ of the models.}
   \label{fig:li}
\end{figure}

Lithium abundances in stars are of great importance 
in astrophysics: they can be used as diagnostics
of stellar evolution, primordial nucleosynthesis as
well as cosmic and Galactic chemical evolution.
Nearly all stellar \element{Li} abundance determinations 
in late-type stars are based on the analysis of the
\ion{Li}{i} resonance line at $6707.8$~{\AA}.
In order to accurately determine \element{Li}
abundances it is therefore necessary to properly model the formation
of this line, taking into account possible effects due
to stellar granulation as well as departures from LTE in general.
Recent investigations of 3D/1.5D non-LTE
\ion{Li}{i} spectral line formation in the Sun 
\citep{kiselman97,kiselman98,uitenbroek98} and metal-poor solar-type stars
\citep{asplund03,barklem03} indicate that non-LTE effects on the derived abundances are significant and
comparable to the ones due to granulation.
In the present study, however, we only explore the effects
of granulation on the formation of the \ion{Li}{i}~$6707.8$~{\AA} line
under the approximation of LTE; 
a detailed 3D non-LTE analysis of \ion{Li}{i} line formation in giant stars is  
deferred to a future paper.
In Table~\ref{tab:li-corr} (Fig.~\ref{fig:li}), 
we present the  3D$-$1D LTE corrections to \element{Li}
abundances derived from the \ion{Li}{i}~$6707.8$~{\AA} line,
assuming a value of $\log{\epsilon(\element{Li})}_\mathrm{1D}=1.0$ 
at all metallicities in the 1D calculations. 
The overall 3D$-$1D LTE \element{Li} abundance
corrections 
as a function of metallicity behave rather similarly to
the ones derived for weak (${\la}40$~m{\AA}) low-excitation lines 
of other neutral elements.
Since, the ionization potential of \ion{Li}{i} is 
just slightly larger than the one of \ion{Na}{i}, the 3D$-$1D LTE
corrections for these two species are typically very similar.

Once again, the differences between the 3D and 1D abundances 
are relatively large and negative (${\la}-0.5$~dex) at very low metallicities,
because of the markedly cooler temperature structures of 3D model atmospheres
compared with the 1D cases, while they become gradually smaller 
as metallicity increases.
Also, as for neutral lines in general, 3D$-$1D LTE \element{Li} abundance
corrections are more pronounced for the series
of model atmospheres with higher $T_\mathrm{eff}$.

\subsection{[\ion{O}{i}] lines}

\begin{figure*}
   \centering
   \resizebox{\hsize}{!}{
   \includegraphics{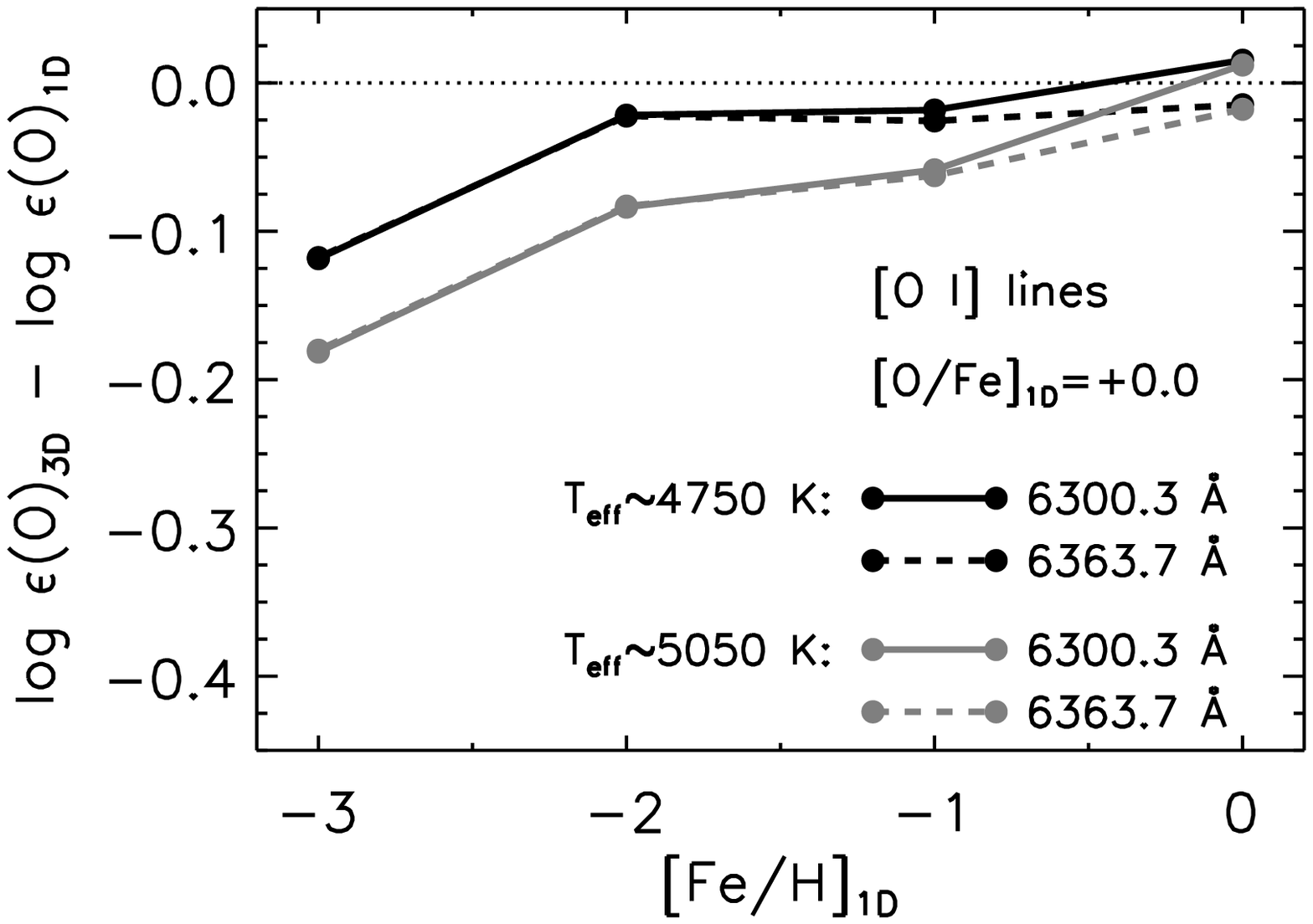} 
   \includegraphics{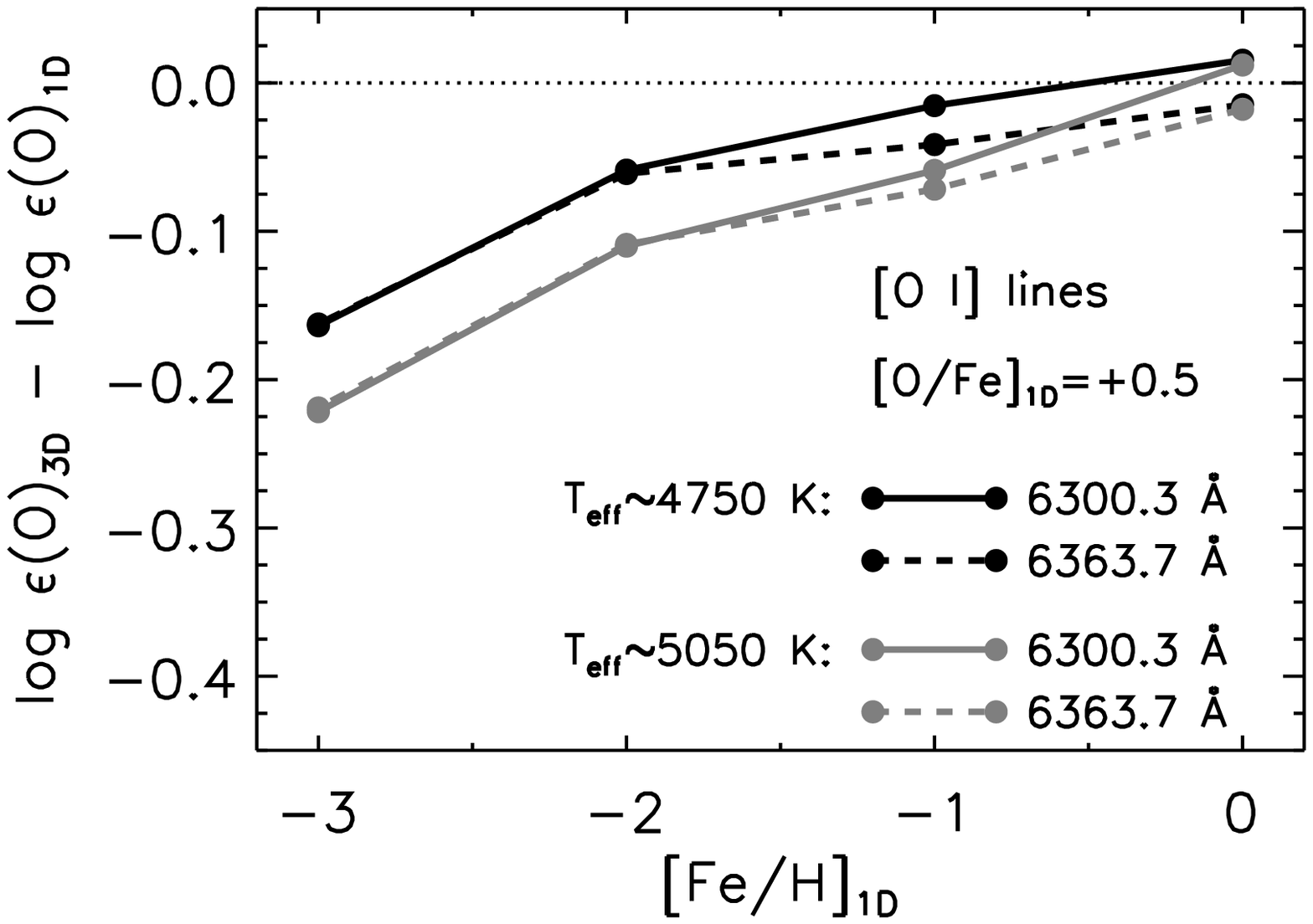} }    
   \caption{3D$-$1D LTE corrections to \element{O} abundances derived
	from the forbidden [\ion{O}{i}] lines at $\lambda=6300.3$~{\AA} and $\lambda=6363.7$~{\AA}
	as a function of metallicity $[\mathrm{Fe/H}]$ of the models.}
   \label{fig:oi}
\end{figure*}

\begin{table}
\caption{Comparison of the 3D and 1D \element{O} abundances derived 
from the [\ion{O}{i}] forbidden lines at $6300.3$~{\AA} and $6363.7$~{\AA}.
The $\log{gf}$ values of the two lines are adopted from \citet{storey00}.
The 3D \element{O} abundances are the ones that reproduce the equivalent
widths computed with 1D models for the tabulated 1D abundances.}
\label{tab:oi-corr}
\centering
\begin{tabular}{cc c c cc}
\hline\hline 
$\langle T_\mathrm{eff}\rangle$&
$[\mathrm{Fe/H}]$ & 
& 
$\log{\epsilon(\element{O})}_\mathrm{1D}$ &

$\log{\epsilon(\element{O})}_\mathrm{3D}$ &
$\log{\epsilon(\element{O})}_\mathrm{3D}$ \\
   
$\mathrm{[K]}$	& & & & $6300.3$~{\AA} & $6363.7$~{\AA} \\ 
\hline
$4697$ &  $+0.0$   &  &  8.89 & 8.90 & 8.88 \\
$4717$ &  $-1.0$   &  &  7.89 & 7.87 & 7.86 \\
$4732$ &  $-2.0$   &  &  6.89 & 6.87 & 6.87 \\
$4858$ &  $-3.0$   &  &  5.89 & 5.77$^{\mathrm{a}}$ & 5.77$^{\mathrm{a}}$ \\
\noalign{\smallskip}
$4983$ &  $+0.0$   &  &  8.89 & 8.89 & 8.87 \\
$5131$ &  $-1.0$   &  &  7.89 & 7.83 & 7.83 \\
$5035$ &  $-2.0$   &  &  6.89 & 6.81 & 6.81 \\
$5128$ &  $-3.0$   &  &  5.89 & 5.71$^{\mathrm{a}}$ & 5.71$^{\mathrm{a}}$ \\
\hline 
\noalign{\smallskip}
$4717$ &  $-1.0$   &  &  8.39 & 8.36 &  8.35 \\
$4732$ &  $-2.0$   &  &  7.39 & 7.33 &  7.33 \\
$4858$ &  $-3.0$   &  &  6.39 & 6.23 &  6.23$^{\mathrm{a}}$ \\
\noalign{\smallskip}
$5131$ &  $-1.0$   &  &  8.39 & 8.33 &  8.32 \\
$5035$ &  $-2.0$   &  &  7.39 & 7.28 &  7.28 \\
$5128$ &  $-3.0$   &  &  6.39 & 6.17 &  6.17$^{\mathrm{a}}$ \\
\hline
\end{tabular}
\begin{list}{}{}
\item[$^{\mathrm{a}}$] The equivalent width of the line is less than $1$~{m\AA}.
\end{list}
\end{table}

The [\ion{O}{i}] forbidden lines at $6300.3$~{\AA} and $6363.7$~{\AA} 
are among the most widely used indicators of \element{O} abundance in cool stars
and halo giants ($T_{\mathrm{eff}}\la5000$~K) in particular 
\citep[e.g.][]{barbuy88,allende01,nissen02,spite05,garciaperez06}
The main advantage concerning the use of [\ion{O}{i}] lines is
that they are expected to form under LTE conditions \citep{kiselman01}; 
the drawback is, on the other hand, that these lines tend to become, in general, very weak
already at $[\mathrm{Fe/H}]\la-2$ in stars with similar surface
gravities as the ones considered here.\footnote{At $[\mathrm{Fe/H}]=-2$ 
and with $[\mathrm{O/Fe}]=0$, the equivalent widths of the two 
[\ion{O}{i}] lines at $6300.3$~{\AA} and $6363.7$~{\AA}
computed using the {\sc marcs} model atmosphere at $T_\mathrm{eff}=4732$~K
are $3.7$~m{\AA} and $1.2$~m{\AA}, respectively.}
The existence of oxygen over-abundances with respect to iron in metal-poor stars 
has been pointed out already by \citet{conti67}.
The exact amount of the over-abundances, however, and their overall trend with
metallicity is still today a matter of debate \citep{garciaperez06}.
In our analysis, we therefore consider, for simplicity, 
two different values of oxygen enhancement with respect
to the scaled solar standard compositions
for the 1D spectral line formation calculations:  
$\mathrm{[O/Fe]}=+0.0$ and $\mathrm{[O/Fe]}=+0.5$ 
(for $\mathrm{[Fe/H]}{\leq}-1$).

In Table~\ref{tab:oi-corr}, we present the differential 3D and 1D \element{O}
abundances as derived from the two [\ion{O}{i}] lines.
The trends of the 3D$-$1D abundance corrections as 
a function of metallicity or effective temperature of the models
(Fig.~\ref{fig:oi}) are qualitatively the same as in the case of
the other neutral species discussed above.
Nonetheless, the 3D$-$1D corrections for [\ion{O}{i}] forbidden lines
are in general significantly smaller.
Because of its high ionization potential ($13.6$~eV), \ion{O}{i}
is, in fact, the dominant ionization stage for oxygen in the line formation
layers of both 3D and 1D models and shows rather little sensitivity
to the differences between the 3D hydrodynamical and 1D hydrostatic 
temperature stratifications at the surface (see also Sect.~\ref{sec:namgca}).
The small differences between the two [\ion{O}{i}] lines
in terms of abundance corrections are due to the different
formation depths of the two features.

\begin{figure*}
   \centering
   \resizebox{\hsize}{!}{
   \includegraphics{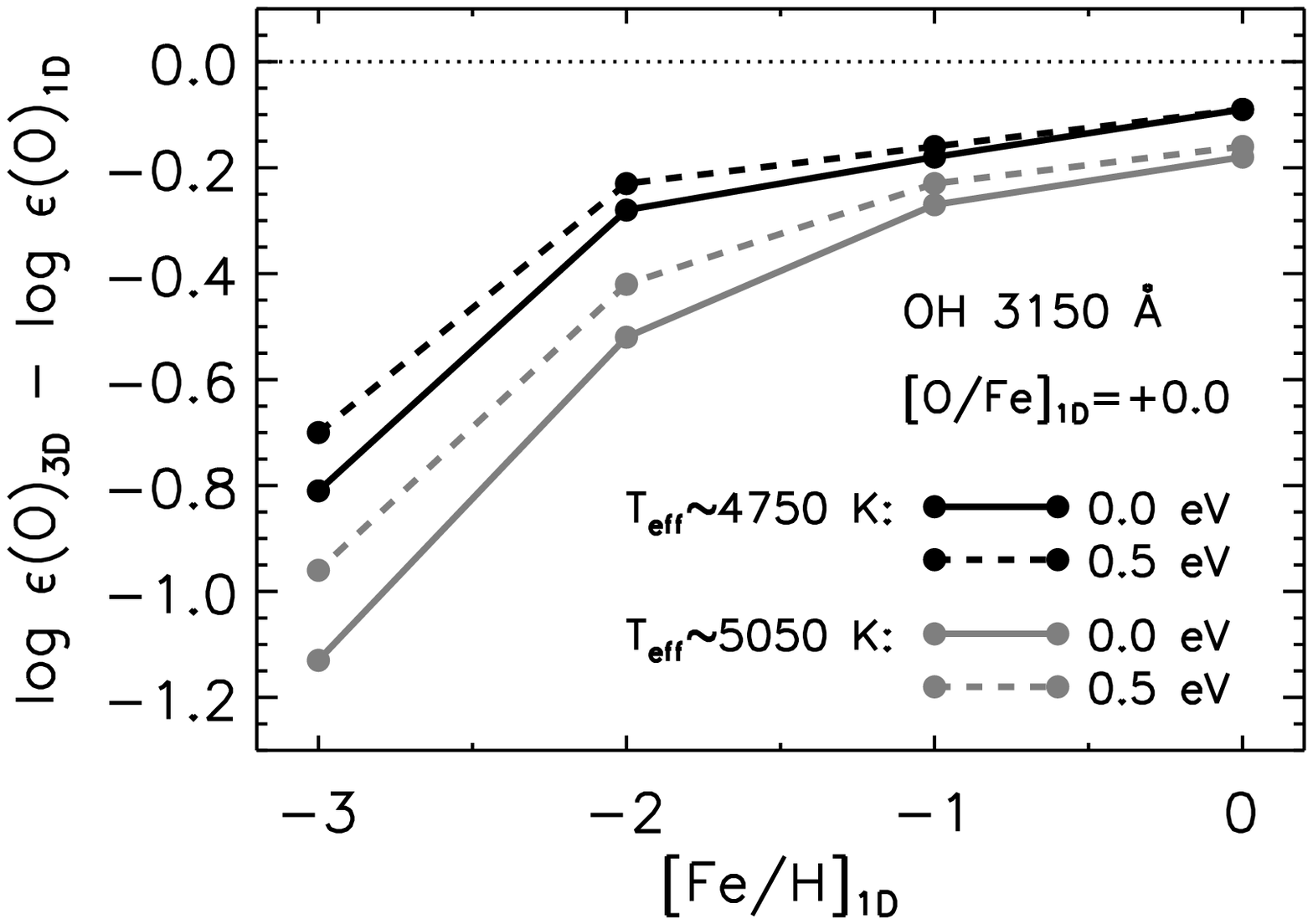} 
   \includegraphics{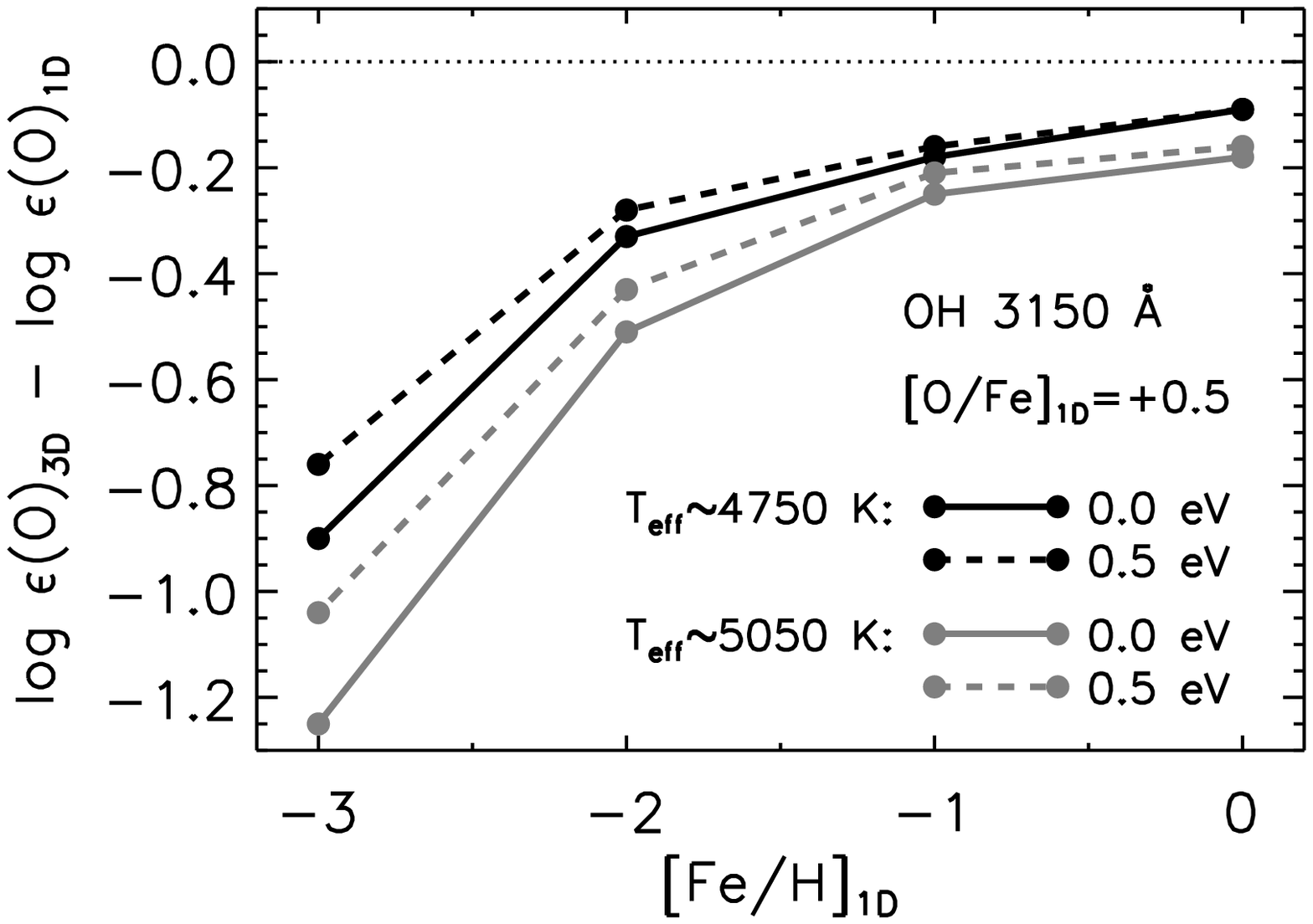} }
   \resizebox{\hsize}{!}{
   \includegraphics{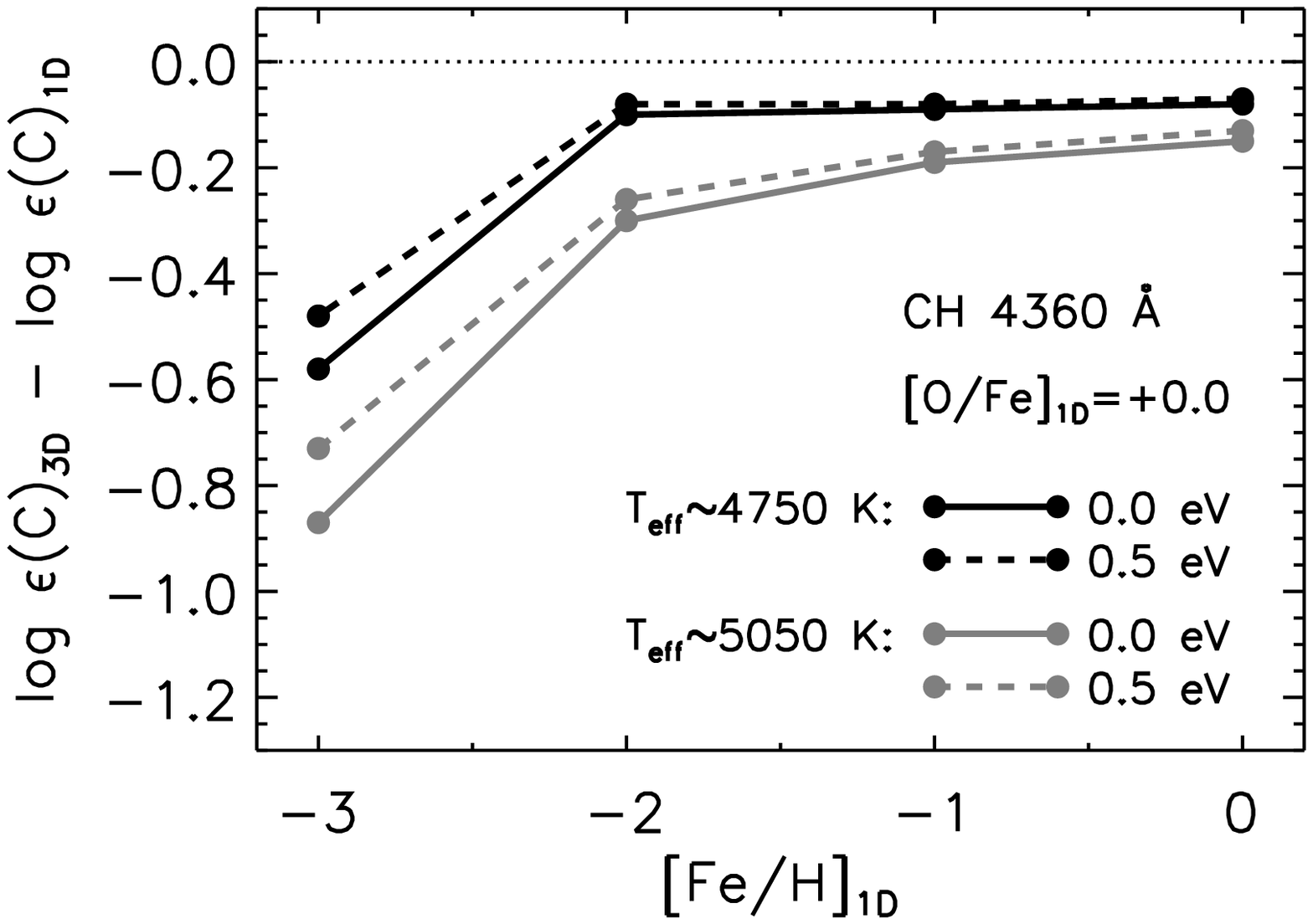} 
   \includegraphics{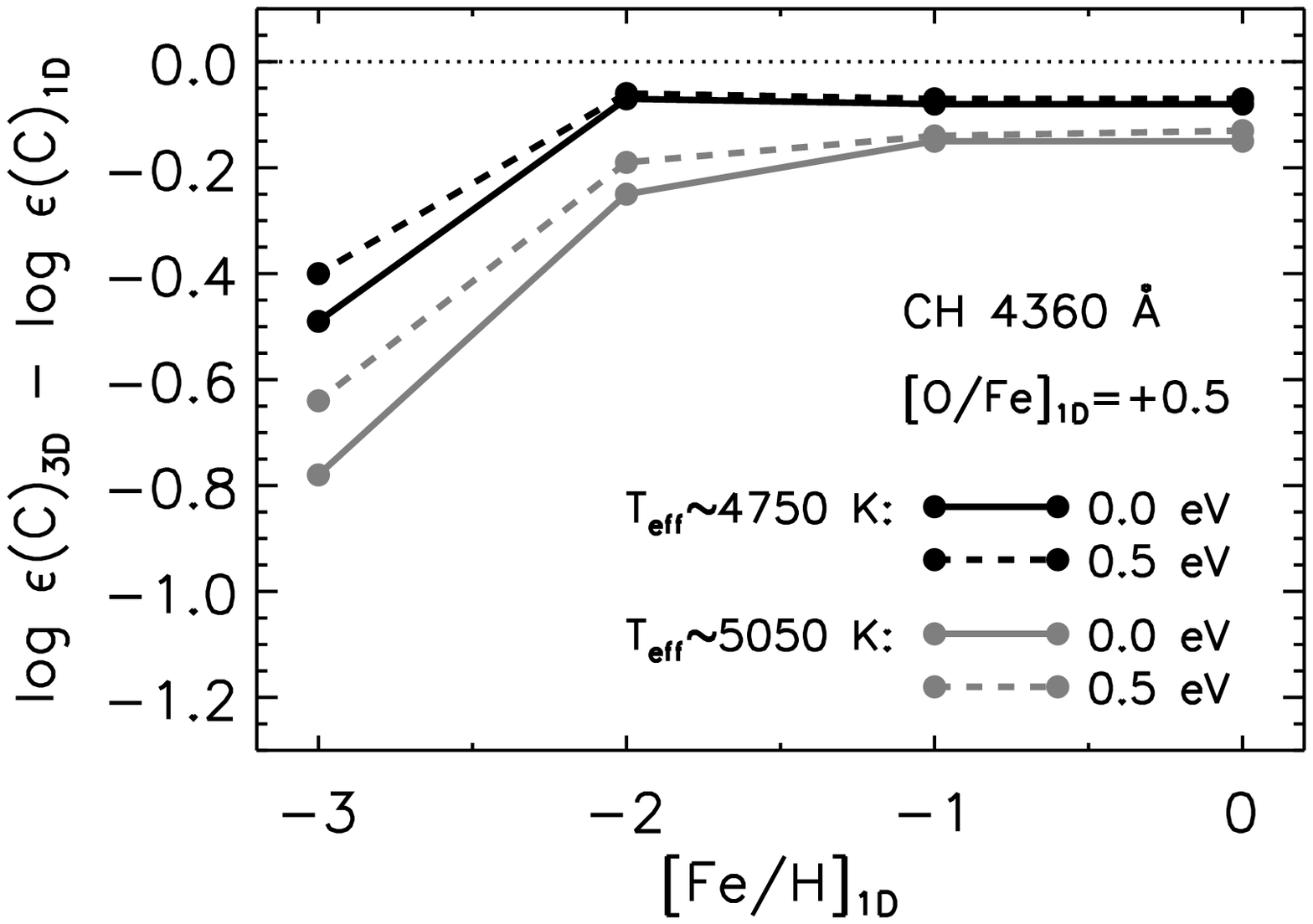} }
   \caption{3D$-$1D LTE corrections to \element{O} and \element{C} abundances derived
	from fictitious OH lines at $\lambda=3150$~{\AA} and CH lines at $\lambda=4360$~{\AA}
	as a function of metallicity $[\mathrm{Fe/H}]$ of the models.}
   \label{fig:oh3150}
\end{figure*}

\begin{table*}
\caption{Differential 3D and 1D \element{C} and \element{O} abundances derived 
from the analysis of weak fictitious CH and OH lines (see Sect.~\ref{sec:spec-form}).
The 3D abundances are the ones that reproduce the equivalent widths of
fictitious lines calculated with 1D models for the tabulated
1D \element{C} and \element{O} abundances. }
\label{tab:choh-corr}
\centering
\begin{tabular}{cc c cc c cc c cc}
\hline\hline 
$\langle T_\mathrm{eff}\rangle$&
$[\mathrm{Fe/H}]$ & 

&

$\log{\epsilon(\element{O})}_\mathrm{1D}$ &
$\log{\epsilon(\element{C})}_\mathrm{1D}$ &

&

$\log{\epsilon(\element{O})}_\mathrm{3D}$ &
$\log{\epsilon(\element{O})}_\mathrm{3D}$ &
&

$\log{\epsilon(\element{C})}_\mathrm{3D}$ &
$\log{\epsilon(\element{C})}_\mathrm{3D}$ \\
   
$\mathrm{[K]}$	& && & && $0$~eV & $0.5$~eV && $0$~eV & $0.5$~eV \\ 
\hline
\noalign{\smallskip}
$4697$ &  $+0.0$    &  &  8.89 & 8.52 &  & 8.80 & 8.80 &  & 8.44 & 8.45 \\
$4717$ &  $-1.0$    &  &  7.89 & 7.52 &  & 7.71 & 7.73 &  & 7.43 & 7.44 \\
$4732$ &  $-2.0$    &  &  6.89 & 6.52 &  & 6.61 & 6.66 &  & 6.42 & 6.44 \\
$4858$ &  $-3.0$    &  &  5.89 & 5.52 &  & 5.08 & 5.19 &  & 4.94 & 5.04 \\
\noalign{\smallskip}
$4983$ &  $+0.0$    &  &  8.89 & 8.52 &  & 8.71 & 8.73 &  & 8.37 & 8.39 \\
$5131$ &  $-1.0$    &  &  7.89 & 7.52 &  & 7.62 & 7.66 &  & 7.33 & 7.35 \\
$5035$ &  $-2.0$    &  &  6.89 & 6.52 &  & 6.37 & 6.47 &  & 6.22 & 6.26 \\
$5128$ &  $-3.0$    &  &  5.89 & 5.52 &  & 4.76 & 4.93 &  & 4.65 & 4.79 \\
\hline 
\noalign{\smallskip}
$4717$ &  $-1.0$    &  &  8.39 & 7.52 &  & 8.21 & 8.23 &  & 7.44 & 7.45\\
$4732$ &  $-2.0$    &  &  7.39 & 6.52 &  & 7.06 & 7.11 &  & 6.45 & 6.46\\
$4858$ &  $-3.0$    &  &  6.39 & 5.52 &  & 5.49 & 5.63 &  & 5.03 & 5.12\\
\noalign{\smallskip}
$5131$ &  $-1.0$    &  &  8.39 & 7.52 &  & 8.14 & 8.18 &  & 7.37 & 7.38\\
$5035$ &  $-2.0$    &  &  7.39 & 6.52 &  & 6.88 & 6.96 &  & 6.27 & 6.33\\
$5128$ &  $-3.0$    &  &  6.39 & 5.52 &  & 5.14 & 5.35 &  & 4.74 & 4.88\\
\hline
\end{tabular}
\end{table*}

\subsection{CH, NH, and OH lines}
Molecule formation shows an extreme and highly non-linear sensitivity
to temperature in the upper layers of late-type stellar photospheres.
Because of this strong temperature dependence, molecular lines are very
susceptible to systematic errors in the thermal structure
of the model atmospheres.
In 3D model atmospheres, the vertical temperature gradients 
are typically steeper in upflows than in downdrafts. 
The temperature contrast reverses in the upper photosphere
some distance above the optical surface, so that 
the gas above the granules becomes cooler than average
in the overshoot region \citep{stein98}.
The temperature contrast in these layers is more pronounced
for the lower metallicity models where the coupling
between the radiation field and the gas is weaker.
Temperatures in the high photospheric layers are generally low;
however, compression due to converging gas flows or shocks
at the boundaries between granules
heat the fluid in the intergranular lanes, raising the
temperature above the radiative equilibrium value.
Consequently, the strengths of molecular lines are expected
to vary across the surface of the model atmospheres, being stronger
in the upflowing regions.
Also, the formation of the spatially and temporally resolved 
molecular line profiles is  biased towards granules
because of their large area coverage, steeper temperature
gradients, and higher continuum intensities.
 
The significantly lower temperatures of the upper
photospheric regions of 3D metal-poor model atmospheres
compared with their 1D counterparts favour 
higher concentrations of CH, NH, and OH molecules for
a given chemical composition, therefore resulting in stronger
molecular lines and negative 3D$-$1D C, N, and O abundance corrections.
Differences between the 3D and 1D predicted molecular line 
strengths are less pronounced at higher metallicities,
in accordance with the similarities between the 1D and mean 3D temperature
stratifications, and are ascribable primarily to
the temperature inhomogeneities in the 3D structures.

\begin{figure}
   \centering
   \includegraphics[width=0.5\textwidth]{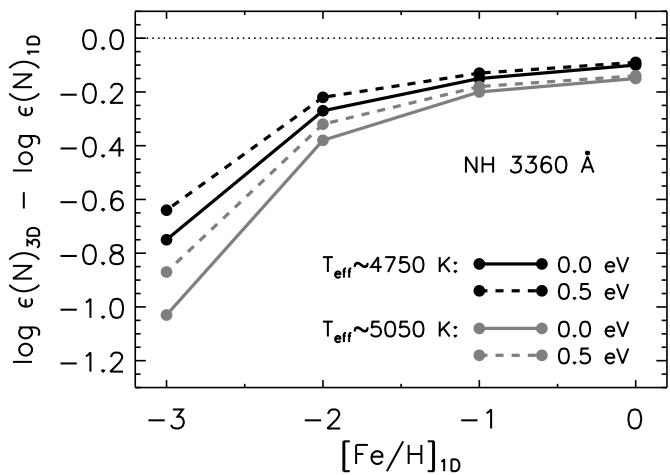} 
   \caption{3D$-$1D LTE corrections to \element{N} abundances derived
	from fictitious NH lines at $\lambda=3360$~{\AA}
	as a function of metallicity $[\mathrm{Fe/H}]$ of the models. }
   \label{fig:nh3360}
\end{figure}

In Tables~\ref{tab:choh-corr} and~\ref{tab:nh-corr} we present
the 3D$-$1D LTE C, N, and O abundance corrections
derived from weak fictitious CH, NH, and OH lines
(with equivalent widths $W_\lambda{\la}30$~m{\AA}), 
as a function of metallicity (see also Fig.~\ref{fig:oh3150} 
and~\ref{fig:nh3360}).
At very low metallicities ($\mathrm{[Fe/H]}\simeq-3$),
the 3D$-$1D corrections are very large and negative
and range between $-0.5$ and $-1.2$~dex 
for the cases here considered.
As expected, the differences between 3D and 1D abundances
become gradually less pronounced at higher metallicities, reducing
to values of approximately $-0.1$~dex at $\mathrm{[Fe/H]}=0$.
The actual magnitude of the abundance corrections for molecular features
depends on several factors such as the species under
consideration,
the line parameters (in particular the lower level excitation potential),
and the effective temperatures of the models.
The 3D$-$1D effects for higher excitation lines at a given metallicity
are generally smaller in accordance with
the sensitivity of these lines to deeper layers where the temperature
differences between 3D and 1D models are less prominent.
The overall larger corrections for hotter models are in agreement
with the larger 3D$-$1D temperature differences in very metal-poor stars
and, at near solar metallicities, with the presence of temperature 
inhomogeneities in the 3D structure and the high non-linearity 
of molecule formation.

The magnitudes of the corrections for CH, NH, and OH 
lines naturally depend on the details of the molecular equilibrium 
with other species, and the various indicators in practice show 
different sensitivities not only to the photospheric temperature structure 
but also to the chemical composition.
For instance, at very low metallicities, 
the corrections for NH and CH lines differ by up to $0.2$~dex
in spite of the similar dissociation energies of the two molecules
($3.42$~eV  and $3.47$~eV, respectively).
Line strengths of CH and OH molecules are also
sensitive to both carbon \emph{and} oxygen atomic abundances.
In Table~\ref{tab:choh-corr} (Fig.~\ref{fig:oh3150})
we compare the results of 3D$-$1D analyses on
OH and CH lines performed assuming, for the 1D calculations,
two different values of oxygen enhancement with respect
to the scaled solar standard compositions:  
$\mathrm{[O/Fe]}=+0.0$ and $\mathrm{[O/Fe]}=+0.5$ 
(for $\mathrm{[Fe/H]}{\leq}-1$), 
with all other metals regularly scaled as usual by the same amount as iron.
Changes in the oxygen abundance alter the number densities of 
of both OH and CH molecules by different amounts in 3D and 1D.
In particular, increasing $\mathrm{[O/Fe]}$ from $+0.0$ to 
$+0.5$ in the 1D calculations results in overall larger 3D$-$1D 
abundance corrections for OH lines and smaller corrections for CH lines.
In practice, however, the effects on the abundance corrections are appreciable only at
very low metallicities.
The trends of 3D$-$1D C and O abundance corrections from CH and OH 
molecular lines at low metallicity depend on the particular choice 
made here for the 1D chemical compositions (with $\mathrm{C/O}<1$) and on the 
sensitivity of molecule formation to temperature in the coolest layers
of 3D models.
In the very metal-poor 1D giant models, most of the carbon and oxygen
are in atomic form, due to the low surface gravity, the overall low C and O
abundances, and the relatively high temperatures of the upper photospheric layers
(${\sim}4000$~K or more). The fraction of carbon and oxygen atoms locked in
CO molecules is negligible and increases only slightly when $\mathrm{[O/Fe]}$
is changed from $0$ to $+0.5$.  
As a result, the 1D photospheric number density of CH molecules also 
remains approximately the same in both the $\mathrm{[O/Fe]}=0$ and 
$\mathrm{[O/Fe]}=+0.5$ cases.
On the contrary, in very metal-poor 3D models, temperatures can be significantly
lower in the upper photosphere and, below ${\sim}3700$~K, a considerable fraction 
of carbon ends up locked in CO.  As $\mathrm{[O/Fe]}$ increases from $0$ to $+0.5$, 
the number density of CO in these cool layers increases further, \emph{reducing} 
the amount of carbon available for the formation of CH molecules. 
Thus, while the number density of CH molecules is still larger than in 1D, 
the 3D$-$1D C abundance corrections become smaller (less negative)
if $\mathrm{[O/Fe]}$ increases.
In the case of oxygen, in the very metal-poor 1D models where the fraction
of CO is negligible, the number density of OH molecules in the photosphere
simply scales proportionally to the O abundance, i.e. it increases by a factor 
of $10^{0.5}{\simeq}3$ when $\mathrm{[O/Fe]}$ is raised from $0$ to $+0.5$.
 In the \emph{coolest} layers of 3D models, where most of the carbon is 
locked in CO, the amount of oxygen available for OH molecule formation 
is also reduced. Assuming $\log(\mathrm{C/O})_{\sun} \simeq -0.3$ 
and that most carbon is in form of CO molecules,
the amount of oxygen available to form OH molecules or remain in atomic form
increases by a factor as high as $(10^{0.8} - 1)/(10^{0.3} - 1)\,{\simeq}\,5\,>\,3$
as  $\mathrm{[O/Fe]}$ is changed from $0$ to $+0.5$; that also explains 
why the 3D$-$1D O abundance corrections derived from [\ion{O}{i}] and OH 
lines become more pronounced as $\mathrm{[O/Fe]}$ increases.

\begin{table}
\caption{Differential 3D and 1D \element{N} abundances derived from the analysis 
of weak fictitious NH lines.
The 3D abundances are the ones that reproduce the equivalent widths of
fictitious lines calculated with 1D models for the tabulated
1D \element{N} abundances.}
\label{tab:nh-corr}
\centering
\begin{tabular}{cc c c cc}
\hline\hline 
$\langle T_\mathrm{eff}\rangle$&
$[\mathrm{Fe/H}]$ & 
& 
$\log{\epsilon(\element{N})}_\mathrm{1D}$ &

$\log{\epsilon(\element{N})}_\mathrm{3D}$ &
$\log{\epsilon(\element{N})}_\mathrm{3D}$ \\
   
$\mathrm{[K]}$	& & & & $0$~eV & $0.5$~eV \\ 
\hline
$4697$ &  $+0.0$   &  & 8.01 & 7.91 & 7.92 \\
$4717$ &  $-1.0$   &  & 7.01 & 6.86 & 6.88 \\
$4732$ &  $-2.0$   &  & 6.01 & 5.74 & 5.79 \\
$4858$ &  $-3.0$   &  & 5.01 & 4.26 & 4.37 \\
\noalign{\smallskip}
$4983$ &  $+0.0$   &  & 8.01 & 7.86 & 7.87 \\
$5131$ &  $-1.0$   &  & 7.01 & 6.81 & 6.83 \\
$5035$ &  $-2.0$   &  & 6.01 & 5.63 & 5.69 \\
$5128$ &  $-3.0$   &  & 5.01 & 3.98 & 4.14 \\
\hline 

\end{tabular}
\end{table}

\section{Restrictions of line formation calculations}
\label{sec:discussion}
\subsection{Scattering}
\label{sec:scattering}
The results of our analysis indicate that the differences
between abundances derived from 3D and 1D model atmospheres
of red giant stars can be considerable, especially at very low
metallicities. 
The presence of large 3D$-$1D corrections can have strong implications
for the interpretation of stellar abundances in terms of Galactic chemical
evolution.
The current 3D analysis, however, still relies on a number of approximations and
assumptions, which can possibly lead to systematic errors
in the derived abundances and must therefore be discussed.

As mentioned in Sect.~\ref{sec:convec-sim} and~\ref{sec:spec-form},
we treat scattering as true absorption 
when solving the radiative transfer equation, both for the convection simulations
and the line formation calculations with 3D and 1D model atmospheres.
This assumption can in general lead to systematic
errors in the predicted temperature stratification of the upper
layers of 3D hydrodynamical model atmospheres where the contribution
of scattering to extinction is significant, and, ultimately,
affect the strengths of synthetic spectral lines.
 Rayleigh scattering of \ion{H}{i} can contribute significantly 
to the total continuous extinction in the UV and blue part
of the spectrum. 
At these wavelengths, implementing scattering as true absorption 
in the detailed line formation calculations underestimates, in general, the
emergent flux in the continuum, resulting in too weak spectral lines \citep{cayrel04}. 
The effect is expected to be more pronounced in very metal-poor stars,
where line-blocking in the continuum is weak, and, in particular,
in metal-poor 3D model atmospheres, where the lower surface temperatures
result in lower electron number densities 
and higher densities of scatterers (\ion{H}{i} particles)
relative to absorbers (namely \element{H}$^-$ particles in the
case of continuous absorption).

In order to quantify the effect of our approximated treatment
of scattering on the photospheric stratifications, 
we have computed a series of 1D {\sc marcs} model atmospheres of
giant stars at various metallicities,
including scattering as true absorption in the solution
of the radiative transfer equation. 
The results of our test indicate that treating scattering as
true absorption leads to \emph{hotter} temperature
stratifications compared with models in which 
scattering is properly taken into account \citep[see also][]{gustafsson75}.
Temperature differences at optical depth $\log{\tau_\mathrm{5000}}=-3$
can reach $300$~K for model stellar atmospheres of very metal-poor
red giants at $\mathrm{[Fe/H]}=-3$; for $\mathrm{[Fe/H]}=-1$ model
atmospheres, the temperature differences at the same optical depth
are smaller and amount to about $100$~K.
On the sole basis of these results, one could be tempted to conclude that the
temperatures at the surface of the metal-poor 3D hydrodynamical models
might be over-estimated.
However, further testing is needed to estimate the magnitude
of the effect on the 3D temperature stratification in metal-poor stars,
and, more importantly, to also assess whether, in this case, it proceeds 
in the same direction.
Either way, the treatment of scattering as true absorption
is likely to significantly affect the balance between radiative heating 
and adiabatic cooling in the upper photospheric layers.

We have also performed some test line formation calculations 
with the average 3D stratifications of our very metal-poor 
model red giant stellar atmospheres and using the
the spectrum synthesis code {\sc bsyn} from the Uppsala package,
which treats continuum scattering properly within the LTE 
approximation.
We find that including scattering as true absorption in the calculations
has relatively little influence on the predicted line strengths, at least for the 
type of giants considered in our study.
Nonetheless, the effect of the above assumption might, in practice, be
larger in full 3D line formation calculations because of the presence
of temperature and density inhomogeneities.
The use of a differential 3D$-$1D abundance analysis ensures, however, that
the uncertainties in the treatment of scattering 
in the spectral line formation calculations
are at least reduced. 

\subsection{Departures from LTE}
\label{sec:nlte}
It is important to emphasize that, in general, spectral lines 
of many of the species considered in our analysis
are expected to suffer from departures from LTE.
In particular,  \ion{Fe}{i} lines are most certainly seriously
affected by non-LTE effects.
The main non-LTE mechanism for \element{Fe} in late-type stellar atmospheres
is over-ionization, which is driven by the radiation field $J_\nu$ in the UV
being larger than the Planck function at the local temperature $B_\nu(T)$.
This causes efficient photo-ionization from \ion{Fe}{i}, leading to
underpopulation of all \ion{Fe}{i} levels compared with LTE
and to \emph{weaker} \ion{Fe}{i} lines.
This, in turn, implies that \element{Fe} abundances derived 
from \ion{Fe}{i} lines are generally \emph{larger} 
in non-LTE than with the assumption of LTE.
At present, there is no consensus on how severe 
the non-LTE effects actually are on \ion{Fe}{i} lines 
in late-type stars \citep[see, e.g., discussion in ][]{asplund05}.
\citet{collet06} have estimated the non-LTE effects 
on \ion{Fe}{i} lines for the extremely metal-poor giant 
HE\,0107$-$5240 by means of a 1D analysis based both on 
a {\sc marcs} model atmosphere and the mean atmospheric stratifications 
inferred from 3D simulations. 
The non-LTE effects there are found to be considerable and opposite
to the 3D$-$1D LTE corrections, confirming that a combined
treatment of both 3D and non-LTE effects is necessary for
accurate \element{Fe} abundance determinations. 
A full 3D non-LTE study of \ion{Fe}{i} spectral line formation
is certainly of high priority, but goes beyond the scope of the
present work and is postponed to a future paper.

Similarly, lines of other species considered in this study,
such as \ion{Li}{i}, \ion{Na}{i}, \ion{Ca}{i}, and \ion{Mg}{i} 
might form out of LTE conditions.
Molecular lines might as well suffer from departures from LTE;
however, non-LTE effects on molecular line formation are, still today, largely
unexplored even in 1D analyses \citep{asplund01}.
In addition, molecule formation could also occur out of LTE:
the rapid cooling occurring at the surface of the 3D simulations
could imply that molecular equilibrium is not reached; 
also, photo-dissociations feeding on the radiation field coming
from deeper layers might move the molecular equilibrium out of LTE.
These aspects are undoubtedly worthy further investigation.

\subsection{Stellar parameters}
The derivation of accurate abundances requires not only 
realistic modelling of the structure of stellar atmospheres and 
of the processes of spectral line formation, but also 
appropriate determination of the fundamental stellar parameters.
In the differential 3D$-$1D analysis presented here,
we have assumed the stellar parameters to be the same
for both 3D and 1D model atmospheres.
However, one could expect the determination of the stellar parameters 
for a given star to depend on whether 1D or 3D models are used 
\citep[see also discussion in ][]{asplund01}.
 For instance, the results of our differential analysis of
\element{Fe} lines suggest that differences in the derived
metallicities cannot indeed be excluded.
Also, differences between the thermal structures
of 3D and 1D model atmospheres can affect ionization equilibria
and, therefore, the determination of spectroscopic surface gravities. 
Finally, temperature inhomogeneities in the continuum-forming layers
of 3D hydrodynamical model atmospheres can, in general,
lead to different emergent flux distributions
compared with the predictions of 1D models and, consequently,
affect effective temperature estimates.

The above discussion on the possible sources of systematic errors 
in the 3D modelling of stellar atmospheres and line formation
can create the impression that the results
presented here are rather uncertain and that analyses
based on 1D model atmospheres are still preferable.
It is important, however, to emphasize that many of 
the uncertainties in the present 3D analysis 
(e.g. departures from non-LTE and molecular equilibrium)
apply as well to the 1D analysis, besides the systematic errors
introduced, in the latter case, by the assumption of hydrostatic equilibrium
and the rudimentary treatment of convective energy transport.

\section{Conclusions}
In the present work, we have investigated the impact of 3D hydrodynamical
model atmospheres of red giant stars on the formation of spectral lines
of various ions and molecules and on the derivation
of elemental abundances.
The differences between the mean 3D temperature stratifications
and corresponding 1D model atmospheres as well as the presence 
of temperature and density inhomogeneities and correlated velocity fields
in the 3D structure, can significantly affect the predicted 
line strengths and, hence, the values of elemental abundances
derived from spectral lines.
Differences between abundance determinations based on 3D and 1D models
are particularly large at low metallicities.
The temperatures in the upper layers of 3D hydrodynamical model
atmospheres of metal-poor giant stars
are in fact significantly lower than the ones predicted by 
classical 1D models computed for the same stellar parameters.
Because of the cooler temperature stratifications of 3D model atmospheres, 
weak lines of neutral species and molecules appear considerably \emph{stronger} in
3D than in 1D, under the assumption of LTE.
The 3D$-$1D corrections for these lines are large and negative.
In particular, we find the 3D$-$1D LTE corrections to CNO abundances
derived from CH, NH, and OH weak low-excitation
to be typically in the range $-0.5$~dex to $-1.0$~dex for
the very metal-poor giants at $[\mathrm{Fe/H}]\simeq-3$ 
considered here.
We also derive large negative corrections to the \element{Fe}
abundances (about $-0.8$~dex) at these metallicities, 
based on the analysis of weak low-excitation Fe~{\sc i} lines. 

In Sect.~\ref{sec:discussion}, we finally discuss possible 
systematic errors affecting the present 3D abundance analyses.
A major source of uncertainty in this respect is
the approximated treatment of scattering as true absorption,
both in the 3D convection simulations and in the line formation
calculations. 
As discussed in Sect.~\ref{sec:scattering}, this approximation
can lead to systematic errors in the estimated temperatures
of the upper layers of 3D model atmospheres as well as in 
the predicted emitted fluxes in the UV, therefore
the strengths of spectral lines.
Also, the assumption of LTE is in general questionable for
many of the ions and molecules considered in the analysis.
In particular, in the case of iron, strong over-ionization feeding on the UV radiation 
field can lead to significantly \emph{weaker} \ion{Fe}{i} lines 
in metal-poor red giants stars compared with the cases where LTE is enforced.
The effects of over-ionization on the strengths of Fe I lines are opposite to the ones of
granulation. 
Departures from LTE should, therefore, accounted for in
the 3D line formation calculations in order to accurately determine 3D$-$1D
corrections to the Fe abundances.

\begin{acknowledgements}
The authors acknowledge support from
the Swedish Foundation for International Cooperation in Research 
and Higher Education and the Australian Research Council.
K. Eriksson and B. Gustafsson are thanked for fruitful discussions.
Finally, the authors would like to thank the anonymous referee
for the very positive and constructive criticism which helped improve
the manuscript.
\end{acknowledgements}

\bibliographystyle{aa} 
\bibliography{rcollet}

\end{document}